\documentclass[12pt,preprint]{article}

\usepackage[utf8]{inputenc}
\usepackage{cancel}
\usepackage{enumitem}
\usepackage{amssymb}
\usepackage{soul}
\usepackage{feynmf}
\usepackage{tabularx}
\usepackage{xcolor}
\usepackage{booktabs}
\usepackage{subcaption} 
\usepackage{colortbl}
\usepackage{authblk}
\usepackage{ulem}
\DeclareUnicodeCharacter{00A0}{ }

\newcolumntype{C}{>{\centering\arraybackslash}X}
\usepackage[T1]{fontenc}
\usepackage{epsfig}
\usepackage{latexsym}
\usepackage{graphicx}
\usepackage{amsmath}
\usepackage{amsfonts}   
\usepackage{amssymb}    
\usepackage{float}
\usepackage{bm}
\usepackage{url}
\usepackage{hyperref} 
\usepackage[nodisplayskipstretch]{setspace}
\def\lsim{\raise0.3ex\hbox{$\;<$\kern-0.75em\raise-1.1ex\hbox{$\sim\;$}}}
\def\gsim{\raise0.3ex\hbox{$\;>$\kern-0.75em\raise-1.1ex\hbox{$\sim\;$}}}

\def    \beq            {\begin{equation}}
\def    \eeq            {\end{equation}}
\def    \bea           {\begin{eqnarray}}
\def    \eea           {\end{eqnarray}}

\def \mn{\mu\nu{\rm SSM}}

\def\g2{{\rm GeV}^2}

\def\sw2{sin^2 \theta_w}

\def\a^tau{\alpha_{\tau}}

\def\beq{\begin{equation}}
\def\eeq{\end{equation}}
\def\beqa{\begin{eqnarray}}
\def\eeqa{\end{eqnarray}}

\def\order#1{\ensuremath{{\cal O}(#1)}}

\newcommand{\tev}{\,\textrm{TeV}}

\newcommand{\newc}{\newcommand}
\newc\BR{BR}
\newc{\akappa}{A_{\kappa} }
\newc\deltagmtwo{\delta (g-2)_{\mu}} 
\newc\deltaamu{\Delta a_{\mu}}

\def\anti{\overline}

\def\la{\lambda}

\newc{\haa}{BR\(h_1\to a_1 a_1\)}
\newc{\abb}{BR\(a_1\to b\anti{b}\)}
\newc{\hbb}{BR\(h_1\to b\anti{b}\)}
\newc{\abund}{\Omega h^2}
\newc\bsgamma{b\rightarrow s \gamma }
\newc\bxsgamma{\overline{B}\rightarrow X_{s}\gamma}
\newc\brbsgamma{\BR(\overline{B}\rightarrow X_s\gamma)}

\allowdisplaybreaks

\usepackage{array, makecell}
\usepackage{boldline}
\usepackage[nosort]{cite}

\title{\bf{
WIMP Dark Matter in the U$\mn$
}}

\author[a,b]{J.~A.~Aguilar--Saavedra\thanks{jaas@ugr.es}}
\author[c,d]{D.~E.~L\'opez-Fogliani\thanks{daniel.lopez@df.uba.ar}}
\author[e,b]{C.~Mu\~noz\thanks{c.munoz@uam.es}}
\author[e,b,f]{M.~Pierre\thanks{mathias.pierre@desy.de}} 

  \affil[a]{Departamento de F\'{\i}sica Te\'orica y del Cosmos,  Universidad de Granada,  E-18071 Granada, Spain}
  \affil[b]{Instituto de F\'{\i}sica Te\'{o}rica (IFT) UAM-CSIC,  Campus de Cantoblanco, 28049 Madrid, Spain}
  \affil[c]{Instituto de F\'isica de Buenos Aires UBA \& CONICET, Departamento de F\'isica, Facultad de Ciencia Exactas y Naturales, Universidad de Buenos Aires, 1428 Buenos Aires, Argentina}
  \affil[d]{Pontificia Universidad Cat\'olica Argentina, 1107 Buenos Aires, Argentina}
    \affil[e]{Departamento de F\'{\i}sica Te\'{o}rica, Universidad Aut\'{o}noma de Madrid (UAM), Campus de Cantoblanco, 28049 Madrid, Spain}
  \affil[f]{Deutsches Elektronen-Synchrotron DESY,  Notkestr. 85, 22607 Hamburg, Germany}

\date{}

\begin{document}

\thispagestyle{empty}
\begin{flushright}
\mbox{}
IFT--UAM/CSIC-21-124\\
DESY-21-181 
\end{flushright}
{\let\newpage\relax\maketitle}

\maketitle

\begin{abstract}
The U$\mu\nu$SSM is a $U(1)'$ extension of the $\mu\nu$SSM supersymmetric model,
where baryon-number-violating operators as well as explicit mass terms are forbidden, and the potential domain wall problem is avoided. 
The gauge anomaly-cancellation conditions impose the presence of exotic quark superfields in the spectrum of U$\mu\nu$SSM models, and allow the presence of several singlet superfields under the standard model gauge group, in addition to the right-handed neutrino superfields. 
The gauge structure implies an additional discrete $Z_2$ symmetry in the superpotential, ensuring the stability of a singlet which behaves as WIMP dark matter without invoking $R$-parity.
We analyze this novel possibility in detail, using the fermionic component of the singlet as the dark matter candidate. In particular, we compute its amount of relic density via $Z'$, Higgs-right sneutrino and dark matter mediated annihilations, and its potential signals in dark matter direct detection experiments. The constraints on the parameter space due to $Z'$ direct searches at the LHC are imposed in the analysis, as well as those from the hadronization inside the detector of the exotic quarks.
Large regions of the parameter space turn out to be in the reach of the upcoming Darwin experiment.
\end{abstract}

\clearpage 

\tableofcontents 

\section{Introduction}
\label{introduction}

One of the crucial evidences for the existence of physics beyond the standard model (SM) is the presence of non-baryonic cold dark matter (DM) in the Universe.
In supersymmetry (SUSY), there are new particles with characteristics {that make them interesting} candidates for DM. In particular, this is the case of weakly interacting massive particles (WIMPs) in $R$-parity conserving SUSY,
such as the neutralino~\cite{Goldberg:1983nd,Ellis:1983wd,Krauss:1983ik,Ellis:1983ew} or the right sneutrino (see Refs.~\cite{Cerdeno:2008ep,Cerdeno:2015ega} and references therein).
{Although these WIMPs have very short lifetimes in $R$-parity violating (RPV) SUSY models, other particles such as the gravitino
or the axino
can nevertheless be valid superWIMP DM candidates. In particular, the lifetimes of the latter turn out to be much longer than the age of the Universe, and interestingly they
produce gamma rays potentially detectable in gamma-ray telescopes.} This was analyzed for the 
gravitino in Refs.~\cite{Borgani:1996ag,Takayama:2000uz,Buchmuller:2007ui,Bertone:2007aw,Ibarra:2007wg,Ishiwata:2008cu,Choi:2010xn,Choi:2010jt,Diaz:2011pc,Restrepo:2011rj,Kolda:2014ppa,Bomark:2014yja} 
in the context of bilinear/trilinear RPV models (for a review, see Ref.~\cite{Barbier:2004ez}),
and in Refs.~\cite{Choi:2009ng,GomezVargas:2011ph,Albert:2014hwa,GomezVargas:2017} in the
`$\mu$ from $\nu$' supersymmetric standard 
model ($\mu\nu$SSM) (for a review, see Ref.~\cite{Lopez-Fogliani:2020gzo}).
Similar analyses for the axino in bilinear/trilinear RPV models were carried out 
in 
Refs.~\cite{Kim:2001sh,Hooper:2004qf,Chun:2006ss,Covi:2009pq,Endo:2013si,Kong:2014gea,Choi:2014tva,Liew:2014gia,Colucci:2015rsa,Bae:2017tqn,Colucci:2018yaq}.
Multicomponent DM scenarios with the axino/gravitino as the lightest supersymmetric particle (LSP) and the gravitino/axino as next-to-LSP (NLSP), 
were also discussed in the $\mn$~\cite{Gomez-Vargas:2019vci,Gomez-Vargas:2019mqk}.

On the other hand, in the recently proposed U$\mn$~\cite{Aguilar-Saavedra:2021qbv,Aguilar-Saavedra:2021smc},\footnote{See also Refs.~\cite{Fidalgo:2011tm,Lozano:2018esg} for similar constructions.} which is a $U(1)'$ extension of the $\mn$~\cite{LopezFogliani:2005yw}, the possible presence of WIMP DM candidates dictated by the anomaly cancellation conditions was proven.\footnote{Several similar constructions based on gauge symmetries broken at the TeV scale with DM candidates arising from gauge anomaly conditions have been already explored outside the context of SUSY in Refs.~\cite{Okada:2012np,DeRomeri:2017oxa,Ballett:2019cqp,Gehrlein:2019iwl,Abada:2021yot,Mandal:2021acg}.}
This is a remarkable result, given that the U$\mn$ is a RPV scenario.  
The aim of this work is to analyze in detail this novel possibility, discussing its cosmological viability as well as its potential signals in DM direct detection experiments.

In the $\mn$,
the presence of RPV couplings involving right-handed (RH) neutrino
superfields, $\hat \nu^c$, solves simultaneously the 
$\mu$ problem~\cite{Kim:1983dt} (for a recent review, see Ref.~\cite{Bae:2019dgg})
of the minimal supersymmetric standard model (MSSM)~\cite{Nilles:1983ge,Haber:1984rc,Martin:1997ns} and the
$\nu$ problem being able to reproduce neutrino data~\cite{LopezFogliani:2005yw,Escudero:2008jg,Ghosh:2008yh,Bartl:2009an,Fidalgo:2009dm,Ghosh:2010zi,Fidalgo:2011tm}.
In the superpotential of this construction, in addition to Yukawa couplings for neutrinos 
$Y^{\nu} \hat H_u\hat L\hat \nu^c$,  the couplings $\lambda\, \hat \nu^c\hat H_d \hat H_u$ are allowed generating an effective $\mu$-term when  the right sneutrinos 
develop electroweak-scale vacuum expectation values (VEVs),
$ \langle {\tilde \nu}_{R}\rangle 
\sim 1$ TeV.

Despite these attractive properties of the $\mn$, there are interesting arguments from the theoretical viewpoint to try to extend the model.
First, we would like to have an explanation for the absence of the  baryon-number-violating couplings 
$\lambda'' \hat u^c\hat d^c\hat d^c$, which together with the lepton-number-violating couplings would give rise
to fast proton decay. Similarly,
the absence of the bilinear terms
$\mu\hat H_u \hat H_d$, $\epsilon \hat H_u \hat L$ and 
${\mathcal M} \hat\nu^c\hat\nu^c$ (and the linear term
$t \hat\nu^c$), which would reintroduce 
the $\mu$ problem and additional naturalness problems, must be explained. 
Finally, since the superpotential of the $\mn$ contains only trilinear couplings, it fetures a discrete $Z_3$ symmetry just like 
the next-to-MSSM (NMSSM)~\cite{Maniatis:2009re,Ellwanger:2009dp}, 
and 
therefore one expects to have also a cosmological domain wall 
problem~\cite{Holdom:1983vk,Ellis:1986mq,Rai:1992xw,Abel:1995wk,Chung:2010cd} unless inflation at the weak scale is invoked.

In Ref.~\cite{Aguilar-Saavedra:2021qbv}, the strategy of considering an extra $U(1)'$ gauge symmetry
to the $\mn$
to explain the absence of the above terms in the superpotential, and to solve the potential domain wall problem, was adopted.
There, the SM gauge group was therefore extended to
$SU(3)\times SU(2)\times U(1)_Y\times U(1)'$.
 Generically, 
with the extra $U(1)'$ one is able to forbid not only the presence of the linear term in the superpotential, but also the above dangerous trilinear and bilinear terms,
since the fields that participate in them can be charged under this group making the terms not invariant under this symmetry.
Besides, the domain wall problem disappears once the discrete symmetry is 
embedded in a gauge 
symmetry~\cite{Lazarides:1982tw,Kibble:1982dd,Barr:1982bb}.

In addition to the above arguments favoring extra $U(1)'$ charges for the fields, this fact also  
avoids the uneasy situation from the theoretical viewpoint of neutrinos being the only fields with no quantum numbers under the gauge group.
For example, in string constructions where extra $U(1)'$ groups arise 
naturally, no ordinary fields appear that are singlets under the full gauge group.
It is worth noting here that explicit 
$SU(3)\times SU(2)\times U(1)_Y\times U(1)'$ four dimensional string models have been built, and they contain typically extra color triplets as well as singlets under the SM gauge group.
See e.g. Refs.~\cite{Casas:1987us,Casas:1988se,Ibanez:1987sn} for models obtained from the compactification of the heterotic string, and Ref.~\cite{Langacker:2008yv} for a review of $Z'$ constructions and references therein.

A crucial characteristic of
the U$\mn$ models built in Ref.~\cite{Aguilar-Saavedra:2021qbv} 
is the presence in their spectrum of exotic matter 
such as extra quark representations or singlets under the SM gauge group, because of anomaly cancellation conditions. The $U(1)'$ charges can make distinctions among these singlets. 
In particular,
some of them behave as RH neutrinos $\hat\nu^c_{i}$, 
 where $i=1,...,n_{\nu^c}$ with $n_{\nu^c}$ the number of RH neutrino superfields,
but others,
$\hat\xi_{\alpha}$, can be candidates for DM because of the $Z_2$ symmetry present in their couplings $k_{i\alpha\beta}\hat\nu^c_{i}\hat\xi_{\alpha}\hat\xi_{\beta}$, 
where $\alpha=1,...,n_{\xi}$ with $n_{\xi}$ the number of $\xi$ superfields.
Since
the scalar and fermionic components of the latter superfields are WIMPs,
the lightest of them can be used as stable WIMP DM.

In Ref.~\cite{Aguilar-Saavedra:2021qbv},
models with three RH neutrinos, $n_{\nu^c}=3$, two DM candidates, $n_{\xi}=2$, and no more extra singlets under the SM gauge group were built. 
Constructions with a different number of these fields, and also including additional singlets were also obtained. {For the sake of {definiteness} and simplicity we will consider here models with $n_{\nu^c}=3$ and $n_{\xi}=2$, but we will also discuss {which} modifications are expected when other singlets are present in the spectrum.}

The paper is organized as follows. Section~\ref{scenarios} will be devoted to the discussion of the U$\mn$ benchmark model we use to illustrate 
the analysis of WIMP DM. 
In Sec.~\ref{sec:DMproduction}, the WIMP DM production via the freeze-out mechanism will be discussed, paying special attention to the analysis of the relic density studying the cross sections corresponding to the annihilation channels.
In Sec.~\ref{constraints}, the constraints on the parameter space of the model from the LHC and DM direct detection experiments, will be explained.
Our results will be shown in Sec.~\ref{results}, where we evaluate the current and potential limits on the parameter space of our DM scenario using the methods described in previous sections.
Finally, our conclusions are left for Sec.~\ref{conclusions}.

\section{The U$\mn$ and dark matter}
\label{scenarios}

Based on the discussion of the Introduction, we consider the following relevant superpotential~\cite{Aguilar-Saavedra:2021qbv}:
\bea
\label{Eq:superpotentialUmunu}
W &=&  
Y^e_{ij} \, \hat H_d\, \hat L_i \, \hat e_j^c
\ +
Y^d_{ij} \, \hat H_d\, \hat Q_i \, \hat d_j^c 
\ -
Y^u_{ij} \, \hat {H_u}\, \hat Q_i \, \hat u_j^c
\ -
Y^\nu_{ij} \, \hat {H_u}\, \hat L_i \, \hat \nu^c_{j}
\nonumber\\
&+&
\lambda_{i} \,  \hat \nu^c_{i}\,  \hat {H_u} \, \hat H_d
\ +
k_{i \alpha \beta}  \, \, \hat \nu^c_{i} \, \hat \xi_{\alpha} \, \hat \xi_{\beta}  
\ +
Y^{\mathbb{K}}_{i j}  \, \hat \nu^c_{i}  \hat{\mathbb{K}}_{j}  \hat{\mathbb{K}}^c_{j}
\,,
\label{superp}
\eea
where the summation convention is implied on repeated indexes, with
$i,j,k=1,2,3$ the usual family indexes of the SM, and $\alpha,\beta=1,2$.
Our convention for the contraction of two $SU(2)$ doublets is e.g.
$\hat {H}_u \,  \hat H_d\equiv  \epsilon_{ab} \hat H^a_u \, \hat H^b_d$,
$a,b=1,2$ and $\epsilon_{ab}$ the totally antisymmetric tensor with $\epsilon_{12}=1$.

Let us remark that a minimum number of exotic quarks is required in U$\mn$ models
by the $[SU(3)]^2 - U(1)'$ anomaly cancellation condition~\cite{Aguilar-Saavedra:2021qbv}.
Namely,
either three pairs of quark singlets of $SU(2)$, ${\mathbb{K}_i}$, or a pair of quark singlets of $SU(2)$,
${\mathbb{K}}$, together with a pair of quark doublets of $SU(2)$, ${\mathbb{D}}$, must be present.
In our analysis, we choose to work with the first solution as shown in the last term of Eq.~(\ref{superp}), but a similar discussion could be carried out with the second solution containing exotic quark doublets ${\mathbb{D}}$.
Note also that the vanishing hypercharge of the RH neutrinos, $y(\nu^c)=0$,
implies that the exotic quarks must have opposite hypercharges to be coupled to each other,
$y({\mathbb{K}}_i^c) = - y({\mathbb{K}}_i)$,
i.e. although in general they are chiral under the $U(1)'$ with their charges $z({\mathbb{K}}_i^c) \neq - z({\mathbb{K}}_i)$, they must be
vector-like pairs under the SM gauge group (like the Higgs doublets $H_u$ and $H_d$).

All the terms in the superpotential of Eq.~(\ref{superp}) are invariant under the $SU(3)\times SU(2)\times U(1)_Y\times U(1)'$ gauge group.
In Table~\ref{table}, we show the quantum numbers of the spectrum of one of the U$\mn$ models built in Ref.~\cite{Aguilar-Saavedra:2021qbv}, which we will use as our benchmark. This model is dubbed {``scenario 1'' in that reference}.
Note nevertheless that the relevant $U(1)'$ charges of $\hat\nu^c_{i}$ and $\hat\xi_{\alpha}$ are common for all models.
{The solution of Table~\ref{table} for the hypercharges of the exotic quarks  
implies that they have the same hypercharges 
as the ordinary quarks. 
Also, the $U(1)'$ charges of the SM matter are leptophobic for this solution since
$z(L)=z(e^c)=0$.}

\begin{table}[t!]
\begin{center}
\begin{tabular}{|c|c|}
\hline
\text{Fields} 
&
$SU(3)\times SU(2)\times U(1)_Y\times U(1)'$
\tabularnewline
 \hline 
 $\hat{Q}_i$ 
&
$\left(3,\  2,\  1/6,\  1/36\right)$
 \tabularnewline
\hline 
 $\hat{u}^c_i$
&
$\left(3,\ 1, -2/3,\ 2/9\right)$
 \tabularnewline
 \hline 
 $\hat{d}^c_i$
 &
$\left(3,\ 1,\ 1/3, -1/36\right)$
 \tabularnewline
 \hline 
 $\hat{L}_i$
&
$\left(1,\ 2,\ -1/2,\ 0\right)$
 \tabularnewline
 \hline 
 $\hat{e}^c_i$
&
$\left(1,\ 1,\ 1,\ 0\right)$
 \tabularnewline
\hline 
$\hat{H}_d$
&
$\left(1,\ 2, -1/2,\ 0\right)$
 \tabularnewline
\hline 
$\hat{H}_u$
&
$\left(1,\ 2,\ 1/2, -1/4\right)$
 \tabularnewline
\hline 
  $\hat\nu^c_{i}$
  &
$(1,\ 1,\ 0,\ 1/4)$   
\tabularnewline
\hline 
 $\hat\xi_{\alpha}$
 &
$(1,\ 1,\ 0, -1/8)$  
\tabularnewline
\hline 
 $\hat{\mathbb{K}}_1 $
&
$\left(3,\ 1, -1/3,\ 1/108\right)$
\tabularnewline
\hline 
 $\hat{\mathbb{K}}_2 $
&
$\left(3,\ 1,\ 2/3,\ -26/108\right)$
\tabularnewline
\hline 
 $\hat{\mathbb{K}}_3 $
&
$\left(3,\ 1,\ 2/3,\ -37/216\right)$
 \tabularnewline
\hline 
$\hat{\mathbb{K}}^c_1 $
&
$\left(\bar 3,\ 1,\ 1/3,\ -28/108\right)$ 
 \tabularnewline
\hline 
$\hat{\mathbb{K}}^c_2 $
&
$\left(\bar 3,\ 1, -2/3,-1/108\right)$ 
 \tabularnewline
\hline 
$\hat{\mathbb{K}}^c_3 $
&
$\left(\bar 3,\ 1, -2/3, -17/216\right)$ 
\tabularnewline
\hline
\end{tabular}
\caption{Chiral superfields 
and their quantum numbers. The fourth entry corresponds to the $U(1)'$ charge of a given field $F$, denoted as $z(F)$ in the text.
}
\label{table}
\end{center}
\end{table}

As mentioned in the Introduction,
a discrete $Z_2$ symmetry is present in the superpotential term of Eq.~(\ref{superp}) containing
the 
superfields of type $\hat\xi_\alpha$,
under which they have a charge $-1$ and the rest of the particle content a charge $+1$.
This symmetry is not an extra requirement but arises from the charge assignment of the model and is a consequence of the gauge anomaly cancellation conditions. Such symmetry remains intact after the spontaneous symmetry breaking of the extra $U(1)^\prime$ by the VEVs of right sneutrinos.
Because of this $Z_2$ symmetry 
the 
superfields of type $\hat\xi_\alpha$
can only appear 
in pairs in the Lagrangian.
As a consequence, 
it is straightforward 
to realize 
that vanishing VEVs for their scalar components,
$\langle\xi_{\alpha}\rangle = 0$, is a solution of the minimization equations. Thus the $Z_2$ symmetry is not broken spontaneously, unless non-renormalizable terms spoil it. The latter depend on the specific construction used. 
We will consider in what follows that the $Z_2$ symmetry is exact, and therefore we will have either the bosonic or the fermionic components of 
$\hat\xi_{\alpha}$ as WIMP DM.

After the minimization of the scalar potential~\cite{Aguilar-Saavedra:2021qbv}, with the choice of CP conservation
the remaining neutral scalars 
\bea
H_d^0 
=
\frac{1}{\sqrt 2} \left(H_{d}^\mathcal{R} + v_d + i\ H_{d}^\mathcal{I}\right),\quad
H^0_u 
=
\frac{1}{\sqrt 2} \left(H_{u}^\mathcal{R}  + v_u +i\ H_{u}^\mathcal{I}\right) ,  
\\
\label{vevu}
\nonumber 
\\
\tilde{\nu}_{iR} 
=
     \frac{1}{\sqrt 2} \left(\tilde{\nu}^{\mathcal{R}}_{iR} + v_{iR} + i\ \tilde{\nu}^{\mathcal{I}}_{iR}\right),
      \quad
  \tilde{\nu}_{iL} 
  =
  \frac{1}{\sqrt 2} \left(\tilde{\nu}_{iL}^\mathcal{R} 
  + v_{iL} +i\ \tilde{\nu}_{iL}^\mathcal{I}\right),
\label{vevnu}
\eea
develop the real VEVs: 
\bea
 \langle H_d^0 \rangle = \frac{v_d}{\sqrt{2}}, \;\;\;\;  \langle H_u^0 \rangle = \frac{v_u}{\sqrt{2}}, \;\;\;\; 
 \langle {\tilde \nu}_{iR}\rangle &=& \frac{v_{iR}}{\sqrt{2}}, \;\;\;\;
 \langle {\tilde \nu}_{iL} \rangle= \frac{v_{iL}}{\sqrt{2}}.
\label{vevs}
\eea

These VEVs are induced by 
the soft SUSY-breaking terms, whose scale is in the ballpark of one TeV.
It is worth noting that whereas $v_{iR}$
are naturally of that order, the VEVs of the left sneutrinos are
$v_{iL}\sim 10^{-4}$ GeV.
These small values arise from their minimization equations because of the proportional contributions to $Y^{\nu}$.
These contributions enter through the F-terms and soft terms in the scalar potential,
and 
are small
due to the 
electroweak-scale seesaw of the model
that
determines $Y^{\nu}\lsim 10^{-6}$.
The smallness of the left sneutrino VEVs for a correct description of the neutrino sector,
compatible with current data, has been shown in Refs.~\cite{LopezFogliani:2005yw,Escudero:2008jg,Ghosh:2008yh,Bartl:2009an,Fidalgo:2009dm,Ghosh:2010zi}. Then, we can define $v^2= v_d^2 + v_u^2 = ({2 m_W}/{g})^2\approx$ (246 GeV)$^2$, where we have neglected the small $\sum_i v^2_{iL}$ contribution to $m_W^2$. 

The first term in the second line of superpotential~(\ref{superp}) generates effectively the
$\mu$-term when the right sneutrinos develop VEVs, 
\bea
\mu=\la_i \frac{v_{iR}}{\sqrt 2}.
\label{muterm}    
\eea
Similarly, the second and third terms generate {the Majorana and Dirac masses} of the fermionic components of the superfields $\hat \xi_{\alpha}$ and $\hat{\mathbb{K}}_{i}$, respectively:
\bea
m_{\tilde \xi_\alpha} = 
2 k_{i\alpha} \frac{v_{iR}}{\sqrt 2}\,,
\;\;\;\;
m_{\mathbb{K}_j}
=\dfrac{Y^{\mathbb{K}}_{ij} v_{iR}}{\sqrt{2}}.
\label{masses}    
\eea

The $Z'$ gauge boson associated to the $U(1)'$, {the neutralino and chargino sectors}, and the neutral Higgs sector are relevant for our analysis of the DM production and annihilation. Let us discuss them in some detail below.

\quad

\begin{figure}[t!]
\begin{center}
\begin{tabular}{ccc}
\includegraphics[height=5.4cm,clip=]{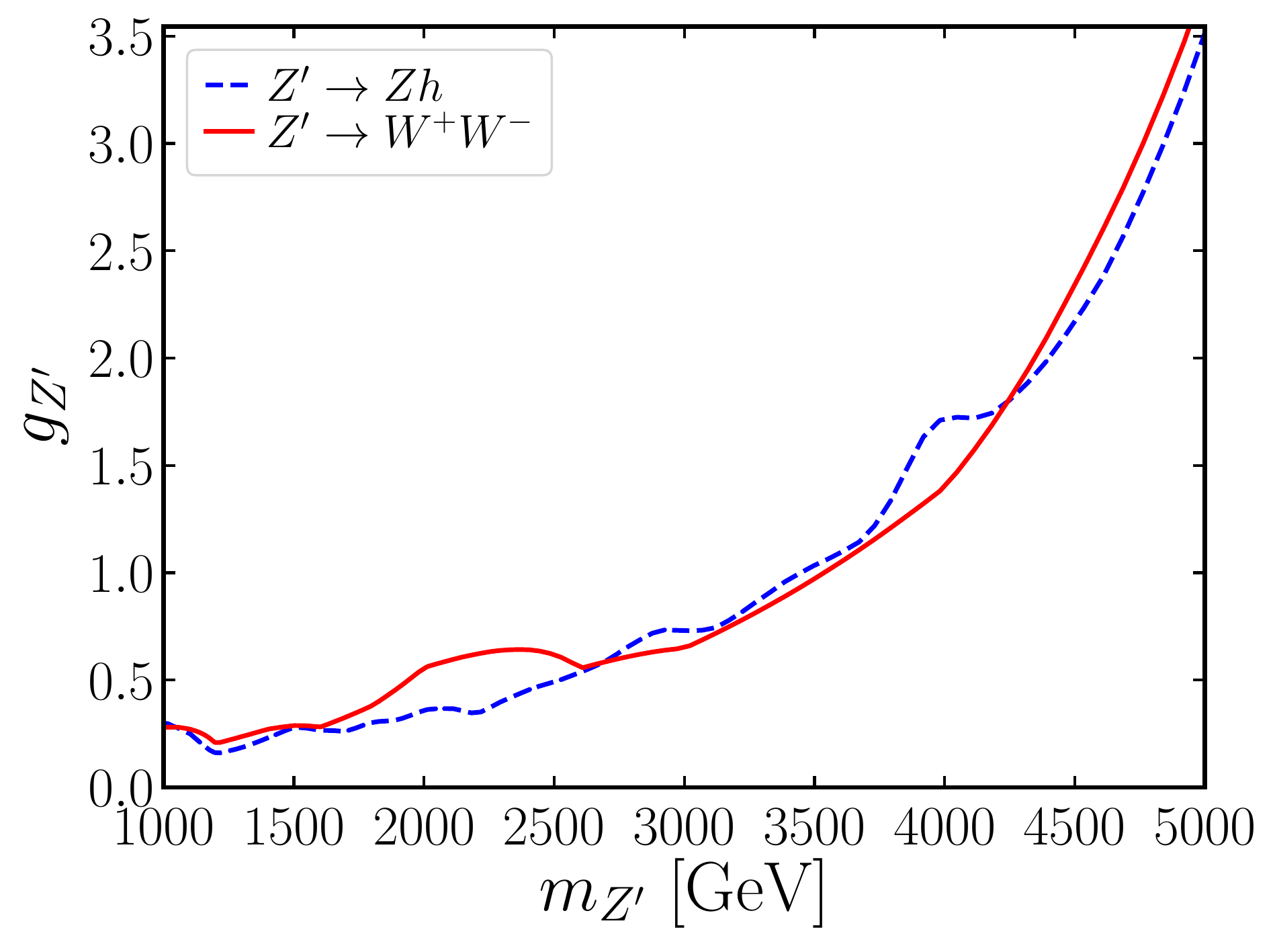} 
\end{tabular}
\caption{Upper Limits for the $U(1)'$ gauge coupling $g_{Z'}$ versus the $Z'$ mass $m_{Z'}$ arising from searches in the final states
$WW$~\cite{Aad:2019fbh} and 
$Zh$~\cite{CMS:2020qrs}, using 
the 
scenario of Table~\ref{table}. 
}
\label{fig:lim1}
\end{center}
\end{figure}

\noindent 
{\it $Z'$ gauge boson}

\noindent
After electroweak breaking the $Z'$ and the SM $Z$ bosons are mixed. This mixing is very small, and in a good approximation one can determine the mass of the $Z'$ boson as~\cite{Aguilar-Saavedra:2021qbv}:
 \bea
m_{Z'}^2 = g_{Z'}^2 \left[ z(H_u)^2 v_u^2 + z(H_d)^2 v_d^2 + 
z(L)^2 v_{iL} v_{iL}
+z(\nu ^c)^2 v_{iR}v_{iR}
\right].
\label{entries}
\eea
where $g_{Z'}$ is the $U(1)'$ gauge coupling.
In this formula, the $U(1)'$ charge $z(F)$ of a field $F$ is shown in Table~\ref{table}.
The limits on $Z'$ masses and mixing from direct searches at the LHC have implications on the VEVs of the singlets that are required to generate the $Z'$ mass. Assuming that all singlet VEVs are of similar order $v_{iR}\sim v_R$, one can write Eq.~(\ref{entries}) as 
\bea
m_{Z'}^2 \approx 3 g_{Z'}^2 z(\nu ^c)^2 
v_{R}^2,
\label{entries22}
\eea
where we have neglected the small contributions from Higgs and left sneutrino VEVs (actually in our
benchmark scenario of Table~\ref{table}, $z(H_d)=z(L)=0$). Taking also into account that the $U(1)'$ charge of the right sneutrinos is $z(\nu^c)= 1/4$, one obtains
$v_R \approx {4 m_{Z'}}/{\sqrt 3 g_{Z'}}$.

Using now results from Ref.~\cite{Aguilar-Saavedra:2021qbv}, we show in Fig.~\ref{fig:lim1} 
the upper limits for the gauge couplings allowed for different $Z'$ masses by direct 
searches at the LHC {for $Z' \to WW$~\cite{Aad:2019fbh}, $Z' \to Zh$~\cite{CMS:2020qrs}, which give the most stringent constraints} in our leptophobic scenario.
These limits were obtained assuming that only $Z'$ decays to SM particles are present, i.e. $Z'\to f\bar f, W^+W^-, Zh$.
Although the presence of new decay modes such as e.g. to sneutrinos or exotic quarks decreases the branching ratio into SM final states, relaxing the current limits of Fig.~\ref{fig:lim1}, this modification is typically small.

Thus, from Fig.~\ref{fig:lim1} one can see e.g. that for masses $m_{Z'} \simeq 1.2, 2.5, 3.3$ TeV, one gets the upper bounds $g_{Z'}\simeq 0.2, 0.4, 0.8$, implying that $v_R\gsim 10$ TeV is necessary to give the $Z'$ boson its mass.
In the numerical analysis of Sec.~\ref{results}, we will scan in 
the range
\begin{equation}
v_R\in [10, 30]\ \text{TeV}.
\label{largevev}
\end{equation}
This result is similar for other leptophobic scenarios with exotic quarks
$\hat{\mathbb{K}}$, $\hat{\mathbb{K}}^c$
and $\hat{\mathbb{D}}, \hat{\mathbb{D}}^c$.
It is worth noting here that one can obtain this VEV hierarchy without increasing an order of magnitude the values of the soft terms. As can be straightforwardly deduced from the minimization equations, it is in fact sufficient to decrease an order of magnitude the $\lambda_i, Y^\nu_i$
couplings, i.e. to values $\lambda_i\sim 0.1$ and $Y^\nu_i\lsim 10^{-8}$. This is because the relevant quantities from the superpotential are the products
$\lambda_i v_{iR}$ and  $Y^\nu_i v_{iR}$.
 and we will adopt the conservative approach of using the values of 
Eq.~(\ref{largevev}).

On the other hand, when other singlet superfields under the SM group are present in the spectrum in addition to right sneutrinos and DM fields, the VEVs of their scalar components contribute to Eq.~(\ref{entries}), and therefore
\bea
m_{Z'}^2 > 3 g_{Z'}^2 z(\nu ^c)^2 
v_{R}^2.
\label{entries222}
\eea
This produces the efect of 
relaxing the limit on the right sneutrino VEV to $v_R\gsim 1$ TeV~\cite{Aguilar-Saavedra:2021qbv}. 
Thus, in order to take into account this scenario in the numerical analysis of Sec.~\ref{results}, we will also scan in 
the range
\begin{equation}
v_R\in [1, 10]\ \text{TeV}.
\label{smallvev}
\end{equation}

{Let us finally remark that non leptophobic models were also built in Ref.~\cite{Aguilar-Saavedra:2021qbv}.
In these cases the LHC limit on the mass of the $Z'$ boson is more stringent, arising from the search of dilepton final states, $Z'\rightarrow \ell\ell$~\cite{Aad:2019fac}. For example, for the mass $m_{Z'}\simeq 5.3$ TeV one gets the uppper bound $g_{Z'}\simeq 0.2$, implying that $v_R\gsim 60$ TeV is necessary to give the $Z'$ boson its mass when no other singlet superfields are present in the spectrum. This is to be compared with the leptophobic lower limit above of $v_R\gsim 10$ TeV.
As we will discuss in Sec.~\ref{numerical}, these high values of the VEV, in particular $v_R > 20$ TeV, are disfavored 
by perturbativity.
On the other hand, when other singlets are present in the spectrum the VEVs of Eq.~(\ref{smallvev}) can be obtained, more naturally if we allow a hierarchy with the VEVs of the other singlets~\cite{Aguilar-Saavedra:2021qbv}. Thus the discussion of Sec.~\ref{numerical} for small VEVs can also be applied to non leptophobic models. 
}

\quad

\noindent 
{\it Chargino and neutralino sectors}

\noindent
Let us focus first our attention on the chargino sector. Only the charged Higgsinos $\tilde H^+_u, \tilde H^-_d$, are relevant for our DM analysis, and they combine to form a 4-component Dirac fermion that we denote as 
$\tilde \chi^{\pm}$.  Its mass is essentially determined by the value of the $\mu$-term:
\bea
m_{\tilde \chi^{\pm}} \approx \mu.
\label{massesneuchar1}    
\eea

In the U$\mu\nu$SSM, the neutralinos, including the extra gaugino $\tilde Z'$, mix with left-handed (LH) and RH neutrinos (and with extra singlinos if present) {because of RPV}~\cite{Aguilar-Saavedra:2021qbv}.
Unlike the $\mn$, in the U$\mu\nu$SSM a cubic term in the superpotential of the type 
$\hat \nu^c \hat \nu^c  \hat \nu^c $ is not allowed by gauge invariance, implying that the RH neutrinos can only acquire large masses through the mixing with the $\tilde Z'$ and the Higgsinos~\cite{Fidalgo:2011tm}.
Then, after diagonalization of the neutralino mass matrix, one obtains that 
one RH neutrino, say $\nu_{1R}$, combines with the LH neutrinos to form four light (three active and one sterile) neutrinos, and that the other two RH neutrinos have EW-scale masses.
Considering for simplicity the mixing between $\tilde{Z}'$ and only one RH neutrino, say $\nu_{3R}$, the two mass eigenvalues (denoting the mass eigenstates as the flavor eigenstates for clarity)
are given respectively by
\bea
m_{\tilde Z',\nu_{3R}} = \frac{1}{2}
\left(\sqrt{{M'_1}^2 + 4 \left(g_{Z'} z(\nu^c) v_{3R} \right)^2} \pm M'_1\right),
\label{eigen}
\eea
where $M'_1$ is the $\tilde{Z}'$ soft SUSY-breaking mass of ${\order{1 \tev}}$.
The neutrino mass can be approximated as $m_{\nu_{3R}}\approx {(g_{Z'}z(\nu^c)v_{3R})^2}/{M'_1}$, and for the ranges of $v_R$ discussed above one obtains $m_{\nu_{3R}}\lsim 1$ TeV for $v_{3R}\sim 30$ TeV 
and $m_{\nu_{3R}} \lsim 100$ GeV for $v_{3R} \sim 1$~TeV.
Using the same strategy for the mixing between the RH neutrino $\nu_{2R}$ and Higgsino, one obtains
\bea
m_{\nu_{2R}} \approx
\dfrac{(\lambda_2 v/\sqrt 2)^2}{\mu}.
\label{massesneu2}    
\eea
However, unlike the result of Eq.~(\ref{eigen}), this formula
is not very accurate, and, in practice, 
$m_{\nu_{2R}}$ varies between that value and an order of magnitude less.
Clearly, this mass is small, $m_{\nu_{2R}}\lsim 1$ GeV, since the experimental bound on chargino masses implies $\mu> 100$ GeV. 
Finally, 
the neutral Higgsino masses are essentially determined by $\mu$ as occurs for the charged Higgsinos:
\bea
m_{\tilde H^0_d, \tilde H^0_u} \approx \mu.
\label{massesneuchar}    
\eea

Let us point out nevertheless, that in 
Ref.~\cite{Aguilar-Saavedra:2021qbv} models 
with additional singlet superfields 
$\hat S$, $\hat N$, 
were built.
These models have
terms in the superpotential such as $\hat S \hat \nu^c  \hat \nu^c $ and
$\hat N \hat S \hat S$, which are useful for reproducing light neutrino masses and mixing angles.
In these cases, the three RH neutrino masses generated by the new scalar VEVs {are naturally 
\bea
m_{\nu_{1,2,3R}} \gsim 1\ \text{TeV}.
\label{rhneutrino}    
\eea
Besides, as discussed above, 
having extra singlets relaxes the limit on the right sneutrino VEV to the range of 
Eq.~(\ref{smallvev}). 
Although for the sake of definiteness we will consider in our numerical analysis of 
Sec.~\ref{results} the RH neutrino masses obtained in Eqs.~(\ref{eigen}) and~(\ref{massesneu2}), we will also discuss the modifications expected when larger masses as in Eq.~(\ref{rhneutrino}) are allowed.

In what follows we will denote the relevant mass eigenstates for our computation discussed 
here, ($\nu_{1R}, \nu_{2R}, \nu_{3R}, \tilde H^0_u, \tilde H^0_d, \tilde Z'$), as neutralinos
$\tilde \chi_i^0$ with $i=1,...,6$.

\quad

\noindent 
{\it Neutral Higgs sector}

\noindent
Because of RPV, Higgses are mixed with right and left sneutrinos. However, the $5\times 5$ Higgs-right sneutrino submatrix is basically decoupled from the
$3\times 3$ left sneutrino submatrix, since the mixing occurs through terms proportional to the small
$Y^\nu_{ij}$ or $v_{iL}$.
{Note that after rotating away the right sneutrino (pseudoscalar Higgs) would be Goldstone boson that generate the $Z'$ ($Z$) mass, we are left with two pseudoscalar right sneutrinos (one pseudoscalar Higgs $A$). For typical values of the parameters, pseudoscalar sneutrinos are heavier than scalar sneutrinos~\cite{Ghosh:2017yeh}, thus we can integrate them out.}
The mass matrix for the relevant scalar eigenstates can be diagonalized via the product of four rotation matrices

\begin{align}
\left( \begin{array}{c}
S_1 \\ S_2 \\ S_3 \\ S_4 \\ S_5
\end{array} \right)
= R_{23}(\theta_1) R_{24}(\theta_2) R_{25}(\theta_3) R_{12}(\alpha) 
\left( \begin{array}{c}
H_d^\mathcal{R} \\ H_u^\mathcal{R} \\
\tilde{\nu}^{\mathcal{R}}_{1R} \\
\tilde{\nu}^{\mathcal{R}}_{2R} \\
\tilde{\nu}^{\mathcal{R}}_{3R} \\
\end{array} \right)
= R_{23}(\theta_1) R_{24}(\theta_2) R_{25}(\theta_3) 
\left( \begin{array}{c}
H \\ h \\
\tilde{\nu}^{\mathcal{R}}_{1R} \\
\tilde{\nu}^{\mathcal{R}}_{2R} \\
\tilde{\nu}^{\mathcal{R}}_{3R} \\
\end{array} \right)
\,,
\end{align}
where $R_{kl}(x)$ is a rotation matrix in the $(k,l)$ plane by an angle $x$ in the usual form, for example with entries $(R_{24}(\theta_2))_{22}=\cos \theta_2$ and
$(R_{24}(\theta_2))_{24}=\sin \theta_2$.

Here $S_{1,...,5}$ denote the mass eigenstates, and $ H_d^\mathcal{R}, H_u^\mathcal{R},
\tilde{\nu}^{\mathcal{R}}_{iR}$ the flavour eigenstates. 
{The singlet components of the SM-like Higgs must be very small because of Higgs experimental data,
$\theta_i\lsim 0.1$.} Thus
$S_{3,4,5} \simeq \tilde{\nu}^{\mathcal{R}}_{1,2,3R}$ is the right sneutrino-like state, $S_2\simeq h$ is the SM-like Higgs state, and {$S_1=H$ is the heavier Higgs state.}
As it is well known, in the so-called decoupling limit {when the pseudoscalar Higgs $A$ 
is much heavier than the $Z$ boson, the $H$ and the charged Higgs $H^{\pm}$ become very heavy and degenerate in mass with $m_H\simeq m_{H^{\pm}}\simeq m_A$.} Besides, the lightest scalar Higgs $h$ and the SM Higgs have very similar properties in agreement with data, with similar couplings to fermions and vector bosons since 
 $ \sin ( \beta - \alpha) \rightarrow 1 $ or equivalently $ \beta - \alpha \simeq \pi / 2 $ (for a review, see~\cite{Djouadi:2005gj}). 
Therefore, working in this limit,
 $H^{\pm}$, $A$, and $H$, can be integrated out in our computation.
 We can work now with the four remaining physical neutral scalars $(S_2, S_{3,4,5}) \simeq 
 (h, \tilde{\nu}^{\mathcal{R}}_{1,2,3R})$ and in the following we denote the mass eigenstates by $(h, \tilde{\nu}^{\mathcal{R}}_{iR})$ for clarity. Changing from the truncated (after diagonalizing the $H_u^\mathcal{R}-H_d^\mathcal{R}$ part and integrating out the heavy $H$) flavour to mass basis can easily be performed in the limit of small mixing angle by the following substitutions:
\begin{equation}
  \tilde{\nu}^{\mathcal{R}}_{iR}  \rightarrow \tilde{\nu}^{\mathcal{R}}_{iR} + \theta_i  h,  \qquad  h  \rightarrow h -  \sum_i \theta_i \tilde{\nu}^{\mathcal{R}}_{iR}\,.
    \label{para}
\end{equation}

\quad

\noindent 
{\bf Dark Matter candidate}

\noindent
In U$\mn$ models, one can have either the bosonic or the fermionic components
of the superfields $\hat\xi_\alpha$ as potentially interesting WIMP DM candidates.
Defining for the bosonic component $\xi_\alpha$ the scalar ($\xi^\mathcal{R}_\alpha$) and pseudoscalar ($\xi^\mathcal{I}_\alpha$) fields as  
\begin{equation}
\xi_\alpha = \frac{1}{\sqrt{2}} ( \xi^\mathcal{R}_\alpha + i \xi^\mathcal{I}_\alpha)\,,
\label{chibosoncomponents}
\end{equation}
their masses squared are given by~\cite{Aguilar-Saavedra:2021qbv}:
\bea
m^2_{\xi^\mathcal{R}_\alpha} 
&=& 
{\frac{1}{2}  g^2_{Z'}\ z(\xi_\alpha)\left[z(H_d)v_d^2 + z(H_u) v_u^2 
+ z(L) v_{iL} v_{iL}
+ z(\nu ^c) v_{iR} v_{iR}
\right] }
\nonumber\\
&+& 
m^2_{\xi_\alpha} + 
m^2_{\tilde \xi_\alpha}
+\,
\left(
\sqrt{2}\ 
 T^{k}_{i\alpha} 
 v_{iR} 
 -
\lambda_i {k_{i\alpha}}
 v_u v_d  
 + Y^{\nu}_{ij} k_{j\alpha}  v_{iL} v_u
 \right)
 \,,
\\
 m^2_{\xi^\mathcal{I}_\alpha} 
&=& 
m^2_{\xi^\mathcal{R}_\alpha} 
- 2
\left(
\sqrt{2}\ 
 T^{k}_{i\alpha} 
 v_{iR} 
 -
\lambda_i {k_{i\alpha}}
 v_u v_d  
 + Y^{\nu}_{ij} k_{j\alpha}  v_{iL} v_u
 \right)
 \,,
\label{eq:chimass2}
\eea
where we are assuming for simplicity that 
the diagonal couplings are dominant for the DM.
In these formulas,
$m_{\xi_\alpha}$ are the soft scalar masses, and $m_{\tilde \xi_\alpha}$ are the masses of the fermionic components $\tilde \xi_\alpha$ in Eq.~(\ref{masses}).
In the case of supergravity, the soft trilinear parameters
$T^k$
are proportional to their corresponding couplings, e.g.
$T^{k}_{11} =A^{k}_{11} k_{11}$ with $A^\kappa$
of {\order{1 \tev}}.

Since the hierarchy $m_{\xi^\mathcal{R}_\alpha} > m_{\xi^\mathcal{I}_\alpha} > m_{\tilde \xi_\alpha}$
can be naturally satisfied, we will use in our analysis the lightest of the fermionic components of the superfields, say $\tilde\xi_{1}$, as the DM particle.
The heaviest state $\tilde\xi_{2}$ can decay for example to
$\tilde\xi_{2}\rightarrow \tilde\xi_{1} \bar{q} q$, and therefore does not play any role in the phenomenology of interest here.
In what follows, we will denote  our DM candidate by $\tilde\xi\equiv \tilde\xi_{1}$.
For some values of the parameters one could have one of the scalar components as the lightest particle, and therefore the DM candidate. This interesting possibility will be discussed in another occasion~\cite{pierre:2020xxx}. \par \medskip

\noindent
In this framework, the relevant contributions from $D$ and $F$-terms to the scalar potential, expressed as a function of flavour eigenstates, read 
\begin{eqnarray}
    V\,&\supset\,&
     \dfrac{g_Z^2}{8} \left( |H_u^0|^2 - |H_d^0|^2\right)^2  +\dfrac{g_{Z^\prime}^2}{32} 
     \left( \sum_i |\tilde{\nu}_{iR}|^2  - |H_u^0|^2 \right)^2 
     \nonumber\\
     &+&
    \lambda^2 \left[ 3 |H_u^0|^2 |H_d^0| ^2 +\left(|H_u^0|^2 + |H_d^0|^2 \right)
    | \sum_i \tilde{\nu}_{iR}  |^2   \right]\,,
    \label{eq:simplified_scalar_pot}
\end{eqnarray}
where $g_Z^2\equiv g^2+g'^2$ with $g$ and $g'$ the $SU(2)$ and $U(1)_Y$ gauge couplings estimated at the $m_Z$ scale by $e=g\sin\theta_W=g'\cos\theta_W$.

In addition, assuming terms larger than the TeV scale, we can integrate out for simplicity sleptons, squarks as well as exotic squarks, and the scalar fields $\xi$. \par \medskip

\bigskip

\noindent
The WIMP DM candidate 
in U$\mn$ models has in general annihilations and interactions with the visible sector generated by $Z'$ and $\tilde\xi$ mediated diagrams or via $\tilde \nu_R - h$ mixing. Their analysis is the aim of the next sections.
Given the large number of parameters, in order to carry out the numerical study we will consider the following flavour-independent parameters denoted by:
\begin{equation}
    \lambda_i\,=\,\lambda \, ,  \qquad   Y^{\mathbb{K}}_{ij}\,=\,Y_{\mathbb{K}} \, ,  \qquad  k_i\,=\, k\,,   \qquad  v_{iR}\,=\, v_R\,,     \qquad  \theta_{i}\,=\, \theta\,.
\end{equation}
In this case, the $\mu$-parameter, the DM mass, and the mass of the exotic quarks 
$m_{\mathbb{K}_i}\equiv m_{\mathbb{K}}$ in Eqs.~(\ref{muterm}) and~(\ref{masses}) are given by:  
\bea
\mu=3\la \frac{v_{R}}{\sqrt 2},
\qquad
m_{\tilde \xi} = 
6 k \frac{v_R}{\sqrt 2}\,,
\qquad
m_{\mathbb{K}}
=3Y_{\mathbb{K}} \frac{v_R}{\sqrt 2},
\label{masses2}    
\eea
 and instead of substitutions~(\ref{para}) we can use
\begin{equation}
  \tilde{\nu}_{iR} 
  \rightarrow  \tilde{\nu}_{iR} + \theta  h, \qquad  h  \rightarrow h - \theta  \sum_{i=1}^3  \tilde{\nu}_{iR}\,,
    \label{para2}
\end{equation}
where we have removed here and in what follows the superscript 
$\mathcal{R}$ for the right sneutrino-like states for clarity of the notation.

\section{Dark matter production}

\label{sec:DMproduction}

For a WIMP DM candidate produced via the freeze-out mechanism, the relic density can be related to the velocity-averaged annihilation cross section {as} $\Omega_{\tilde \xi} h^2 \propto 1/\langle \sigma v \rangle$. For values of the cross section $\langle \sigma v \rangle \simeq 3 \times 10^{-26}~\text{cm}^3~\text{s}^{-1}$~\cite{Steigman:2012nb,Arcadi:2017kky,Mambrini}, the expected relic abundance matches the most recent measurement by the Planck collaboration $\Omega_{\tilde \xi} h^2 =0.11933 \pm 0.00091$~\cite{Aghanim:2018eyx}. To estimate the cross section we perform the usual expansion in terms of powers of the mean DM velocity $\bar{v}_{\tilde \xi}$ evaluated at the DM freeze-out temperature $m_{\tilde \xi}/T_\text{F}\simeq 20$~\cite{Gondolo:1990dk,Jungman:1995df}, which is only valid away from poles or kinematic thresholds~\cite{Griest:1990kh}. The relevant diagrams contributing to DM annihilation are depicted in Fig.~\ref{fig:diagrams}. Given the complexity of the model, a reliable estimate of the total DM density can only be determined numerically. However for completeness, we 
provide in the following analytical expressions for the cross sections corresponding to the main annihilation channels.

\begin{figure}
\centering

\includegraphics[width=\linewidth]{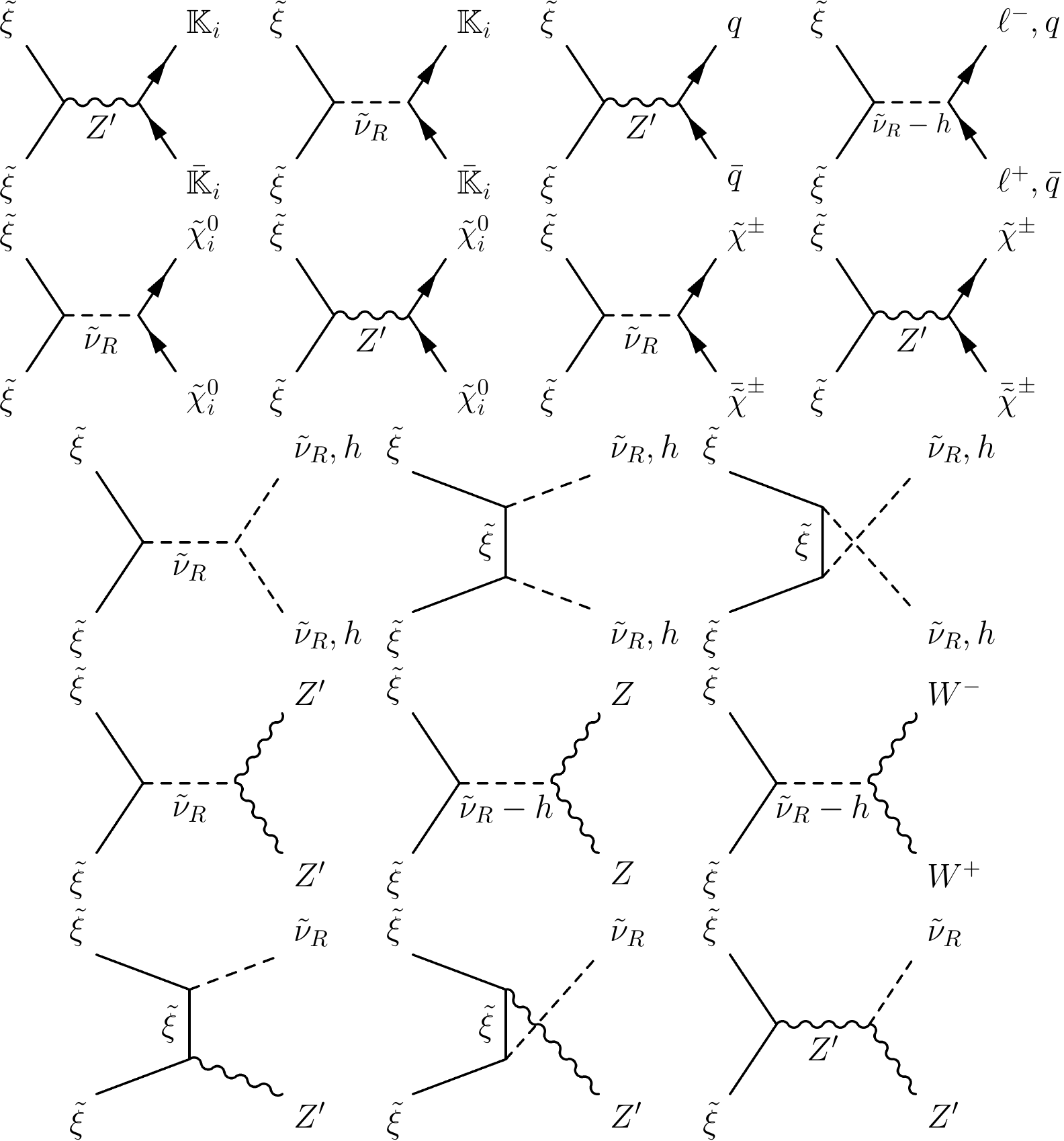}
\vspace{0.5cm}
\caption{Relevant diagrams contributing to DM annihilation.}
\label{fig:diagrams}
\end{figure}

\paragraph{Annihilation to exotic quarks: $ \tilde \xi \tilde\xi \rightarrow \bar{\mathbb{K}}\mathbb{K}.$} DM annihilation to exotic quarks is shown in the first two diagrams of the first line of Fig.~\ref{fig:diagrams}. 
The velocity expansion of the annihilation cross section gives a gauge induced term which is $s$-wave dominated and a $p$-wave term induced by $s-$channel $\tilde\nu_R$ mediated diagrams:
\begin{equation}
    \langle \sigma v_{\tilde \xi} \rangle_{\bar{\mathbb{K}} \mathbb{K}} =  \sum_{i=1}^3   \langle \sigma v_{\tilde \xi} \rangle_{\bar{\mathbb{K}}_i \mathbb{K}_i} \simeq  \frac{9 g_{Z^\prime}^4 m_{\mathbb{K}}^2 }{16384 \pi  m_{Z^\prime}^4} \left(1-\dfrac{m_{\mathbb{K}}^2}{m_{\tilde \xi}^2}\right)^{1/2}
    + \bar{v}_{\tilde \xi}^2  \frac{m_\mathbb{K}^2 m_{\tilde\xi }^4 }{ 4\pi v_R^4  
    \left(m_{\tilde{\nu}_R}^2-4 m_{\tilde\xi }^2\right)^2} \left(1-\frac{m_\mathbb{K}^2}
    {m_{\tilde\xi }^2}\right)^{3/2}
    \label{eq:sigmav_KK}
\end{equation}
where the large factor in the denominator of the first term arises from the $U(1)'$ charges of the fields (a similar comment applies to other annihilation contributions below). The leading contribution, mediated by the $Z^\prime$, is not velocity suppressed but proportional to the squared masses of the exotic quarks in the final state. At $\mathcal{O}(\bar{v}_{\tilde \xi}^2)$, terms are induced by $s-$channel exchange of the right sneutrinos. Although additional terms induced by gauge interactions should also be present, they are not displayed since they would be negligible compared to the $s$-wave dominant term. In the case where the mass of the $Z^\prime$ is basically generated by the VEV of the right sneutrino, i.e. no extra singlets are present, we have 
$m_{Z^\prime}^2\approx 3 g_{Z^\prime}^2 v_R^2/16$ (see Eq.~(\ref{entries22})) and in the limit $m_{\tilde{\nu}_R} \ll m_{\tilde\xi }$  and $m_{\mathbb{K}} \ll m_{ \tilde \xi}$ the two terms
in Eq.~(\ref{eq:sigmav_KK}) become independent on both the gauge coupling and DM mass, and scale precisely in the same way, up to the velocity suppression for the $\tilde \nu_R$-mediated contribution, as
\begin{equation}
    \langle \sigma v_{\tilde \xi} \rangle_{\bar{\mathbb{K}} \mathbb{K}}  \simeq  \frac{ m_{\mathbb{K}}^2 }{64 \pi v_R^4} \Big( 1+ \bar{v}_{\tilde \xi}^2 \Big)\, \simeq 3 \times 10^{-26} \text{cm}^3 \text{s}^{-1} \left( \dfrac{m_{\mathbb{K}}}{3\,\text{TeV}}\right)^2 \, \left( \dfrac{2.5\,\text{TeV}}{v_R}\right)^4\,.
     \label{eq:sigmav_KK_simp}
\end{equation}
In the case  $m_{\tilde{\nu}_R} \gg m_{\tilde\xi }$, the second term is even further suppressed compared to the first one, and therefore we conclude that the gauge contribution always dominates. However, if the $Z^\prime$ mass receives additional contributions from singlets, {we have 
$m_{Z^\prime}^2 > 3 g_{Z^\prime}^2 v_R^2/16$ (see Eq.~(\ref{entries222})), and} the velocity-suppressed term could become as important or larger than the gauge contribution.

\paragraph{Annihilation to quarks: $ \tilde \xi  \tilde \xi \rightarrow \bar{q} q.$}
This is shown in the last two diagrams of the first line of Fig.~\ref{fig:diagrams}. 
The DM annihilation cross section to a pair of quarks gives in the non-relativistic limit
\begin{equation}
    \langle \sigma v_{\tilde \xi} \rangle_{\bar{q}q} \simeq \frac{3 A_q^2 m_q^2  z(\xi)^2 g_{Z'}^4 }{4 \pi  m_{Z^\prime}^4} \left(1-\frac{m_q^2}{m_{\tilde \xi }^2}\right)^{1/2} + \bar{v}_{\tilde \xi}^2 \frac{3 \theta^2 m_q^2 m_{\tilde \xi}^4  \left(m_h^2-m_{\tilde \nu_R}^2\right)^2 }{4 \pi v^2 v_R^2 \big(m_h^2-4 m_{\tilde \xi}^2\big)^2 \big(m_{\tilde \nu_R}^2-4 m_{\tilde \xi}^2\big)^2} \left(1-\frac{m_q^2}{m_{\tilde \xi}^2}\right)^{3/2}\,
    \label{eq:sigmav_qq}
\end{equation}
where $A_\psi\equiv(z(\psi_L)-z(\psi_R))/2$ is the axial $U(1)^\prime$ charge of a generic 4-component fermion $\psi$, with $z(\psi_R)\equiv-z(\psi^c)$ in the notation of Table \ref{table}. {As for Eq.~(\ref{eq:sigmav_KK}), the cross section is $s-$wave dominated and proportional to the squared masses of the outgoing fermionic states.}
The term $\mathcal{O}(\bar{v}_{\tilde \xi}^2)$ is induced by mixing via $s-$channel mediation of the RH sneutrinos $\tilde \nu_R$ and Higgs-like state $h$. Using for the gauge contribution the annihilation to top quarks in the final states, one obtains
\begin{equation}
    \langle \sigma v_{\tilde \xi} \rangle_{\bar{t}t}  \, \simeq \,  2.3 \times 10^{-26} \,  \left( \dfrac{400~\text{GeV}}{v_R} \right)^{4}  \text{cm}^3\, \text{s}^{-1}\,,
\end{equation}
which requires a small value for the VEV $v_R$ to achieve the correct relic density. {For the mixing term, one obtains}
\begin{equation}
    \langle \sigma v_{\tilde \xi} \rangle_{\bar{t}t}  \,\simeq \, 4.5 \times 10^{-27} \, 
    \theta^2 \left( \dfrac{1000~\text{GeV}}{v_R} \right)^{2}  \text{cm}^3\, \text{s}^{-1}\,,
\end{equation}
which can hardly dominate over the first term of Eq.~(\ref{eq:sigmav_qq}), since $\theta$ is typically small. 
{Clearly, the second term can dominate only close to resonances $m_{h}\simeq 2 m_{\tilde \xi}$ or $m_{\tilde \nu_R}\simeq 2m_{\tilde \xi}$.}

\paragraph{Annihilation to leptons: $ \tilde \xi  \tilde \xi \rightarrow \ell^+ \ell^-.$} Given the fact that leptons are uncharged under the new $U(1)^\prime$ symmetry, as compared to $\bar q  q$ annihilations, $Z^\prime$-mediated diagrams are no longer present but diagrams induced by scalar mixing $\theta$ are still present. Therefore, the analytical dependence of the cross section for $\ell^+ \ell^-$ would be essentially be the same as the second term of Eq. (\ref{eq:sigmav_qq}), which is suppressed by the masses of the fermionic final states. For this reason we expect annihilations to leptons to be subdominant $\langle \sigma v_{\tilde \xi} \rangle_{\ell^+ \ell^-} \ll \langle \sigma v_{\tilde \xi} \rangle_{\bar{t}t}$ and therefore will be discarded in the following.

\paragraph{Annihilation to neutralinos: $\tilde \xi\tilde \xi \rightarrow \tilde \chi_i^0 \tilde \chi_i^0.$}
DM annihilation to neutralinos is shown in the first two diagrams of the second line of Fig.~\ref{fig:diagrams}. 
The contributions from $Z^\prime$ mediated diagram are
\begin{equation}
    \langle \sigma v_{\tilde \xi} \rangle_{\tilde \chi_i^0\tilde \chi_i^0}  \simeq  \dfrac{g_{Z^\prime}^4 m_{\tilde \chi_i^0}^2 }{8192 \pi m_{Z^\prime}^4} \left(1-\frac{m_{\tilde \chi_i^0}^2}{m_{\tilde \xi}^2}\right)^{1/2} \simeq  2.8 \times 10^{-26}   g_{Z^\prime}^{4}
    \left(\dfrac{m_{\tilde \chi_i^0}}{200~\text{GeV}} \right)^{2} 
    \left(\dfrac{160~\text{GeV}}{m_{Z^\prime}} \right)^{4}  \text{cm}^3 \text{s}^{-1} \,,
    \label{eq:annihilation_neutralinos}
\end{equation}
for $i=1,2,3,4$. There are also diagrams mediated by $\tilde \nu_{iR}$ that give a contribution
\begin{equation}
   \langle \sigma v_{\tilde \xi} \rangle_{\tilde \chi_i^0\tilde \chi_6^0} \, \simeq \, \bar v_{\tilde \xi}^2  \frac{g_{Z^\prime}^4 k^2 m_{\tilde \xi}^2 }{32 \pi \big(m_{\tilde \nu_R}^2-4 m_{\tilde \xi}^2\big)^2}\,,
\end{equation}
for $i=1,2,3$. In addition there are annihilations to a Higgsino pair 
\begin{equation}
   \langle \sigma v_{\tilde \xi} \rangle_{\tilde \chi_4^0\tilde \chi_5^0} \, \simeq \, \bar v_{\tilde \xi}^2  \frac{\lambda^2 m_{\tilde \xi}^4 }{8 \pi v_R^2 \big(m_{\tilde \nu_R}^2-4 m_{\tilde \xi}^2\big)^2}\,.
\end{equation}

\paragraph{Annihilation to charginos: $\tilde \xi \tilde \xi \rightarrow \bar{\tilde \chi}^\pm  \tilde \chi^\pm.$} These are the last two diagrams shown in the second line of Fig.~\ref{fig:diagrams}. The cross section can be expressed as:
\begin{equation}
   \langle \sigma v_{\tilde \xi} \rangle_{\bar{\tilde \chi}^\pm \tilde \chi^\pm} \, \simeq \,  \dfrac{g_{Z^\prime}^4 m_{\tilde \chi^\pm}^2 }{16384 \pi m_{Z^\prime}^4 } \left(1-\frac{m_{\tilde \chi^\pm}^2}{m_{\tilde \xi}^2}\right)^{1/2} + \bar v_{\tilde \xi}^2  \frac{ 9 k^2 \lambda^2 m_{\tilde \xi}^2 }{4 \pi  \big(m_{\tilde \nu_R}^2-4 m_{\tilde \xi}^2\big)^2} \left(1-\frac{m_{\tilde \chi^\pm}^2}{m_{\tilde \xi}^2}\right)^{3/2} \,,
\end{equation}
where additional terms $\mathcal{O}(\bar v_{\tilde \xi}^2 )$ induced by gauge interactions should also be present but are subdominant.

\paragraph{Annihilation to scalars: $ \tilde \xi \tilde \xi \rightarrow \tilde{\nu}_R \tilde{\nu}_R,  \tilde{\nu}_Rh, h h.$} The diagrams of DM annihilation to scalars are shown in the third line of Fig.~\ref{fig:diagrams}. The various contributions can be decomposed as
\begin{equation}
    \langle \sigma v_{\tilde \xi} \rangle_\text{scalars} \,=\,   \langle \sigma v_{\tilde \xi} \rangle_{\tilde \nu_{R} \tilde \nu_{R} } + \langle \sigma v_{\tilde \xi} \rangle_{\tilde \nu_{R}  h } +
    \langle \sigma v_{\tilde \xi} \rangle_{ h h  } \, ,
\end{equation}
where at leading order in $\theta \ll 1$ we have
\begin{equation}
   \langle \sigma v_{\tilde \xi} \rangle_{\tilde \nu_{R} \tilde \nu_{R} } \, \equiv \, 3 \langle \sigma v_{\tilde \xi} \rangle_{\tilde \nu_{iR} \tilde \nu_{iR} }+ 3 \langle \sigma v_{\tilde \xi} \rangle_{\tilde \nu_{iR} \tilde \nu_{jR} }  \, \simeq \, \bar{v}_{\tilde \xi}^2 \frac{\big(9 g_{Z^\prime}^4 v_R^4+320 g_{Z^\prime}^2 m_{\tilde \xi}^2 v_R^2+3072 m_{\tilde \xi}^4\big)}{589824 \pi m_{\tilde \xi}^2 v_R^4}\,,
\end{equation}
\begin{equation}
    \langle \sigma v_{\tilde \xi} \rangle_{\tilde \nu_{R}  h }\,\equiv \, 3 \langle \sigma v_{\tilde \xi} \rangle_{\tilde \nu_{iR}  h }\,\simeq \,\bar{v}_{\tilde \xi}^2 \frac{\left(g_{Z^\prime}^2 v_R-16 \lambda^2  (2 v+3 v_R)\right)^2}{393216  \pi m_{\tilde \xi}^2 v_R^2}\,,
\end{equation}
\begin{equation}
   \langle \sigma v_{\tilde \xi} \rangle_{h h }\,\equiv \,   \bar{v}_{\tilde \xi}^2 \frac{ \left(g_{Z^\prime}^2-48 \lambda ^2\right)^2}{262144 \pi m_{\tilde \xi}^2 }\,.
\end{equation}
In the limit where $g_{Z^\prime}, \lambda \rightarrow 0$ and $m_{\tilde{\nu }_R} \ll m_{\tilde\xi}$ we obtain
\begin{equation}
     \langle \sigma v_{\tilde \xi} \rangle_\text{scalars} \,\simeq \,   \bar{v}_{\tilde \xi}^2 \frac{m_{\tilde \xi }^2 }{192 \pi v_R^4} \, \simeq \, 3.8 \times 10^{-26} \,  \left( \dfrac{m_{\tilde \xi}}{3~\text{TeV}} \right)^{2}  \left( \dfrac{700~\text{GeV}}{v_R} \right)^{4}  \text{cm}^3\, \text{s}^{-1}  \, .
     \label{eq:sigmavscalars}
\end{equation}

\paragraph{Annihilation to gauge bosons: $\tilde \xi  \tilde \xi \rightarrow Z^\prime  Z^\prime, Z  Z, W^+  W^-.$} The diagrams are shown in the fourth line of Fig.~\ref{fig:diagrams}.
The DM annihilation cross section to a pair of $Z^\prime$ gives in the non-relativistic limit
\begin{equation}
    \langle \sigma v_{\tilde \xi} \rangle_{Z^\prime Z^\prime} \simeq \frac{ g_{Z^\prime}^4 m_{\tilde \xi }^2}{32768 \pi  \big(m_{Z'}^2-2 m_{\tilde \xi }^2\big)^2}\left(1-\frac{m_{Z'}^2}{m_{\tilde \xi }^2}\right)^{5/2}\,.
    \label{eq:sigmav_ZpZp}
\end{equation}
Annihilations to electroweak gauge bosons are given by
\begin{equation}
    \langle \sigma v_{\tilde \xi} \rangle_{W^+ W^-} \,  \simeq \, \bar{v}_{\tilde \xi}^2 \frac{e^4  \theta^2 v^2 m_{\tilde \xi}^2 m_{\tilde \nu_R}^4 }{512 \pi c_W^4 s_W^4 m_Z^4  \big(m_{\tilde \nu_R}^2-4 m_{\tilde \xi}^2\big)^2 v_R^2}  \, \simeq \, 2 \langle \sigma v_{\tilde \xi} \rangle_{ZZ}\,.
\end{equation}

\paragraph{Annihilation to right sneutrino and $Z'$ gauge boson: $\tilde \xi  \tilde \xi \rightarrow \tilde \nu_R Z^\prime.$} The diagrams are shown in the fifth line of Fig.~\ref{fig:diagrams}.
The DM annihilation cross section in the non-relativistic limit gives
\begin{equation}
    \langle \sigma v_{\tilde \xi} \rangle_{\tilde \nu_R Z^\prime} =3 \langle \sigma v_{\tilde \xi} \rangle_{\tilde \nu_{Ri} Z^\prime} \, \simeq \, \frac{k^2}{16 \pi v_R^2} \, \simeq \, 2.6 \times 10^{-26} \, \text{cm}^3\, \text{s}^{-1} k^2 \left( \dfrac{3~\text{TeV}}{v_R} \right)^{2} \,,
    \label{eq:sigmavZpnur}
\end{equation}
{where we are using $m_{Z^\prime}\ll m_{\tilde\xi}$ and  Eq.~(\ref{entries22}).}

\section{Current bounds}
\label{constraints}

\subsection{Constraints from the LHC}
\label{constraintslhc}

We already discussed in Sec.~\ref{scenarios} the constraints on the parameter space of the model due to $Z'$ direct searches at the LHC, and we refer to the reader to that section. 
Let us then discuss in this section the effects of the presence
of exotic quarks/squarks in the spectrum of the U$\mn$, dictated by the anomaly-cancellation conditions.
This type of particles can be produced at the LHC, and 
it is sensible to assume that they will hadronize
inside the detector into color-singlet states, known in the literature as R-hadrons. Thus, bound states of exotic {quarks/squarks} combined with SM quarks can be produced at the LHC (for a review, see e.g. Ref.~\cite{Kang:2007ib}). 
If the R-hadrons have a lifetime that implies stability on collider timescales, 
the current lower bounds at the LHC on their exotic constituents are of about 1.2 TeV~\cite{Aaboud:2019trc}. 

\subsection{Constraints from cosmology}
\label{constraintscosmo}

The presence of the heavy quarks discussed above, with the charge content specified in Table \ref{table}, implies a cosmological history that would require a detailed investigation. A discrete $Z_2$ symmetry arising from the anomaly-cancellation conditions is present in their superpotential term of Eq.~(\ref{superp}), similarly to the case of the DM fields.
Other terms involving these exotic quarks such as gauge interaction terms with the $Z^\prime$, gluinos, gauginos or terms in the scalar potential also feature this $Z_2$ symmetry. This implies that either such quarks or their scalar partners are stable depending on the mass hierarchy.
Since they are both electrically charged and also charged under QCD, the present abundance of R-hadrons is constrained.

As summarized in Ref.~\cite{Kudo:2001ie}, experimental searches inside sea water of stable charged massive particles (CHAMPs) $X$, forming anomalous water molecules $HXO$, seem to imply the present day bound
$n_{X} /n_H \lsim 10^{-14}$ for masses $m_X\gsim 1~\textrm{TeV}$~\cite{Smith:1982qu}, where
$n_{X}$ and $n_H$ correspond to the number densities of $X$ particles and Hydrogen atoms in the Earth, respectively. 
In~\cite{Kudo:2001ie} the authors carry out an estimate of how many $X$ particles are expected inside the sea water, depending on whether they are in the halo or in the galactic disk, with the result
\begin{equation}
  \left(\dfrac{n_X}{n_H} \right)_{\text{Earth}}\,\simeq \, (3-6) \times 10^{-5} 
  \left(\dfrac{\text{GeV}}{m_X} \right) \Omega_X h^2,
     \label{eq:Yamaguchiabundance0}
\end{equation}
where several assumptions have to be made, such as that the local fraction of the $X$'s energy density relative to that to the DM (baryons) in the halo nearby the Earth (galactic disk) traces its global fraction in the whole Universe.
Transferring this result to our case of R-hadrons, one straightforwardly obtains
\begin{equation}
  \left(\dfrac{n_\text{R}}{n_H} \right)_{\text{Earth}}\,\simeq \, 
  10^{-14} \left(\dfrac{Y_\text{R}}{10^{-18}} \right),
     \label{eq:Yamaguchiabundance}
\end{equation}
where the yield $Y_\text{R} \equiv n_\text{R}/s$, with $s$ being the entropy density.

To get an idea of the value of the yield, let us
assume that the reheating temperature is large enough such that the exotic quarks
$\mathbb{K}$
are produced and thermalized with the rest of the SM particles in the early universe. For a temperature $T>m_\mathbb{K}$, these particles are essentially relativistic and abundant with a density $n_\mathbb{K}\sim T^3$. Once the temperature drops below $ T \lesssim m_\mathbb{K}$, these particles become non-relativistic, their density exponentially suppressed by a factor $e^{-m_\mathbb{K}/T}$ and kinetic equilibrium is maintained by processes such as $\bar{\mathbb{K}} \mathbb{K} \rightarrow \bar q q $ mediated by QCD interactions. Around $ T \lesssim m_\mathbb{K}/25$, the Hubble expansion rate becomes larger than the interaction rate and these particles decouple from the thermal bath giving a yield $Y_\mathbb{K} 
\sim 10^{-14}$,
using an order of magnitude of perturbative QCD estimate. However, during the freeze-out process, Sommerfeld enhancement in the annihilation process\footnote{generated by diagrams involving several gluon exchange between the initial state legs.} $\bar{\mathbb{K}} \mathbb{K}  \rightarrow \bar q q $, as well as the formation of bound states containing particles $\mathbb{K}$, affect the perturbative prediction by decreasing the expected final yield by one order of magnitude~\cite{Kang:2006yd,Mitridate:2017izz,DeLuca:2018mzn,Gross:2018zha}. Below the QCD confinement temperature $T_\text{QCD}\sim \Lambda_\text{QCD}\simeq~180~\text{MeV}$, the exotic quarks would form 
R-hadrons
with masses expected of order $m_\text{R}\sim m_\mathbb{K}$.
These hadrons can collide and form bound states with typically a large angular momentum,
and relax progressively to states of lower angular momentum 
by emitting pions or photons as the temperature decreases giving typically $Y_\text{R}\sim 10^{-18}$ for 
$m_\text{R} \sim$ 10 TeV~\cite{Kang:2006yd,DeLuca:2018mzn,Gross:2018zha}. This gives an upper bound on the possible cosmological abundance for these particles, as any additional annihilation channel\footnote{such as diagrams mediated by $\tilde \nu_R$.} would contribute positively to the annihilation cross section, and results in an additionally depletion of their abundances. 

Thus, from Eq.~(\ref{eq:Yamaguchiabundance}) 
one obtains that the presence of stable R-hadrons is allowed, although close to the experimental bound discussed above.
On the other hand, deriving this bound relies on the assumption that the exotic hadrons accumulate in sea water. 
As pointed out in the most recent analysis of Ref.~\cite{DeLuca:2018mzn}, 
testing a sample of sea water does not necessarily lead to bounds, because the atoms that contain heavy hadrons sink to the bottom.
The authors also discuss the compatibility with experimental bounds of other unusual signals of these strongly interacting massive particles (SIMPs). 
Similarly to the previous case, they argue that the Earth once was liquid, so that the primordial heavy hadrons sank to the core of the Earth, undergoing $\bar{\mathbb{K}} \mathbb{K}$ annihilations. Therefore, in order to set bounds, they consider the smaller secondary abundance of SIMPs, because the Earth captures all primordial SIMPs which still are in galactic clouds, encountered along its trajectory. However, the capture cross sections of SIMP by nuclei are very uncertain. If they are not captured and sunk, their present density in the crust is negligible small. If SIMPs get captured in nuclei, assuming that their capture cross sections are similar to the measured capture cross sections of neutrons by nuclei, then their local density is compatible with bounds for a cosmological abundance $10^5$ times smaller than DM.
This abundance occurs precisely for a yield $Y_\text{R}\sim 10^{-18}$.
A similar conclusion is obtained for SIMP searches performed in meteorites~\cite{DeLuca:2018mzn}.

\vspace{0.25cm}

\noindent
Given the above discussions, we will only impose the LHC lower bound of 1.2 TeV on the masses of R-hadrons.

\subsection{Constraints from dark matter direct detection}

DM direct detection experiments are already excluding regions predicted by theoretical models, by analyzing the elastic scattering on target nuclei through nuclear recoils. In this section we will study the predictions of our DM scenario for the spin-independent and spin-dependent scattering cross sections, and compare them with the current and upcoming experimental constraints.

\bigskip

\noindent
{\bf DM-nucleon spin independent cross section}

\noindent
{As can be deduced from the diagram of DM annihilation to quarks mediated by the Higgs-portal induced by scalar mixing of Fig.~\ref{fig:diagrams}, spin-independent (SI) interactions can be mediated similarly. This is shown in the left diagram of Fig.~\ref{fig:diagram_DD}.
An interesting aspect of our DM scenario is that exotic quarks also contribute to direct detection signals via the mediation by right sneutrinos, as shown in the middle diagram of Fig.~\ref{fig:diagram_DD}. As the presence of these particles is required by anomaly cancellation conditions, their contribution is a rather general prediction of the U$\mu \nu$SSM.}

\begin{figure}
\centering

\includegraphics[width=\linewidth]{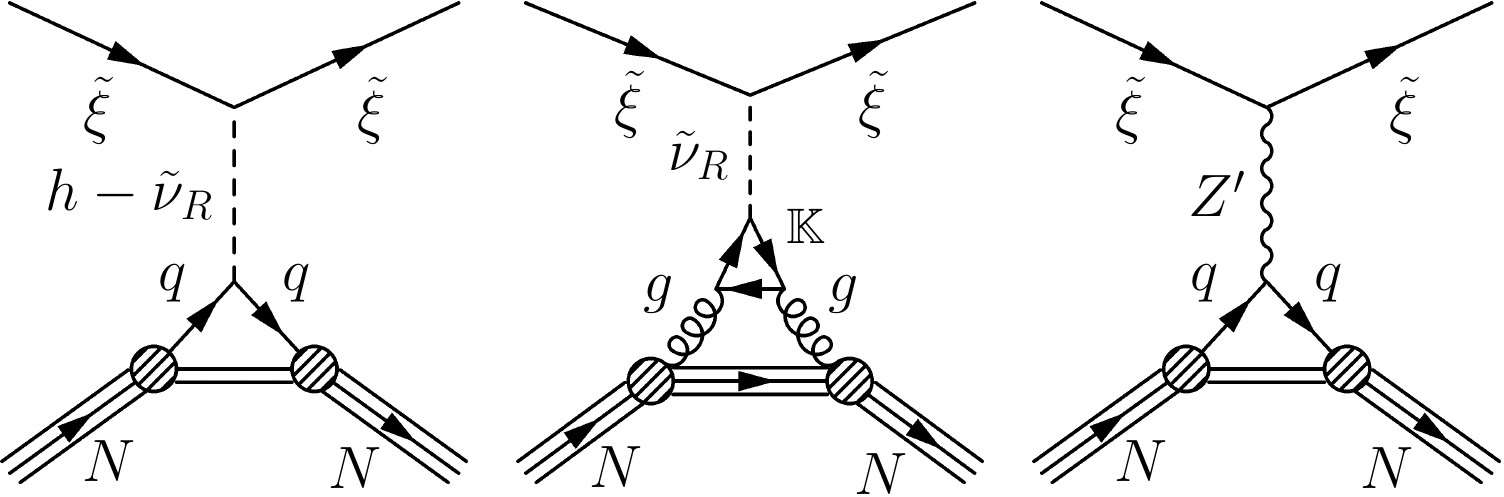}
\caption{Diagrams contributing to DM scattering with nucleons mediated 
by (left panel) SM-like Higgs $h$ and right sneutrinos $\tilde \nu_R$ via mixing, 
(middle panel) right sneutrinos $\tilde \nu_R$ and induced by loop effects involving heavy exotic quarks $\mathbb{K}$, (right panel) $Z^\prime$ gauge boson.}
\label{fig:diagram_DD}
\end{figure}

After integrating out scalar mediators, the following 4-point operators are generated at the nuclear scale
\begin{equation}
\mathcal{O}_{\tilde \xi}^q \,= \,  c_{\tilde \xi}^{q} \bar{\tilde \xi} \tilde  \xi \Bar{q} q
\,,  \qquad    
    \mathcal{O}_{\tilde \xi}^\mathbb{K} \,=  \, \sum_{i=1}^3 c_{\tilde \xi}^{\mathbb{K}_i} \Bar{\mathbb{K}}_i \mathbb{K}_i \bar{\tilde \xi} \tilde  \xi \,,
\end{equation}
where $q$ denotes a generic SM quark, and the Wilson coefficients are
\begin{equation}
c_{\tilde \xi}^{q}\,= \, \dfrac{\theta m_q m_{\tilde \xi} }{2 v_R v}\left( \dfrac{1}{m_h^2}-\dfrac{1}{m_{\tilde \nu_{R}}^2}\right) \,,\qquad 
c_{\tilde \xi}^{\mathbb{K}_i}\,=\,\dfrac{m_{\mathbb{K}_i} m_{\tilde \xi}}{6 v_R^2 m_{\tilde{\nu}_R}^2} 
\,,
\end{equation}
giving rise to the following DM-nucleon SI operator
\begin{equation}
    \mathcal{O}_{\tilde \xi}^N = C_N^{{\tilde \xi}} \bar{\tilde \xi} \tilde \xi \bar{N} N \,.
\end{equation}
The Wilson coefficient $ C_N^{{\tilde \xi}}$ can be expressed as a sum over contributions from SM quarks and exotic quarks, as~\cite{Cirelli:2013ufw} 
\begin{equation}
    C_N^{\tilde \xi}= \sum_{q=u,d,s}c_{\tilde \xi}^q \dfrac{m_N}{m_q} f_{Tq}^{(N)}+\dfrac{2}{27} \sum_{q=c,b,t}c_{\tilde \xi}^q \dfrac{m_N}{m_q} f_{TG}^{(N)} +\dfrac{2}{27} \sum_{i=1}^3c_{\tilde \xi}^{\mathbb{K}_i} \dfrac{m_N}{m_{\mathbb{K}_i}} f_{TG}^{(N)} \,,
\end{equation}
where $f_{Tq}^{(N)}$ and $f_{TG}^{(N)}$ are the contributions of the quark $q$ and the gluons to the nucleon mass~\cite{Cirelli:2013ufw}. The second and third terms in the previous equation represent contributions from heavy SM quarks and exotic quarks, respectively, that have been integrated out. In our case, contributions from SM quarks are suppressed by the mixing angle between the right sneutrinos and the SM-like Higgs.

Summing all contributions, the total DM-nucleon SI scattering cross section can be expressed as:
\begin{equation}
    \sigma_N^\text{SI}\,=\, \dfrac{ 4 \mu_{ \tilde \xi N}^2  m_{\tilde \xi}^2 m_N^2 }{\pi v_R^2 }  \left[
     \dfrac{\theta  }{2  v} \left( \dfrac{1}{m_h^2}-\dfrac{1}{m_{\tilde \nu_{R}}^2}\right) \left( \sum_{q=u,d,s}  f_{Tq}^{(N)}     +  \dfrac{6}{27} f_{TG}^{(N)} \right) 
  +   \dfrac{1}{27 v_R m_{\tilde{\nu}_R}^2} f_{TG}^{(N)} 
     \right]^2 \,.
    \label{crossSI}
\end{equation}
{where $\mu_{ \tilde \xi N}\equiv m_{\tilde \xi} m_N /(m_{\tilde \xi} +m_N) $ is the reduced mass of the $\tilde \xi-N$ system.}
It is worth noting that
even though the second term of the cross section is induced by the exotic quarks, it does not depend explicitly on their masses. This is a counter-example of the intuitive idea that heavy particles do decouple in the limit where their masses become infinite\footnote{However, this picture is limited by the fact that for a given value of $v_R$, increasing the exotic quark masses corresponds to increasing the coupling $Y_{\mathbb{K}}$ which would go beyond perturbativity at some point and this analysis would not be valid anymore.}. It is due to the fact that the coupling between exotic quarks and right sneutrinos mediating the scattering cross section, is proportional to the exotic quark masses. This factor cancels out the mass suppression from the exotic quark propagators in the middle diagram of Fig.~\ref{fig:diagram_DD}. A similar effect occurs in the SM in the loop-induced decay of the Higgs boson into a gluon pair, when the mass of top quark tends to infinity.

{In the limit $m_{\tilde{\nu}_R}\rightarrow\infty$, the $\theta$-dependent first term
of Eq.~(\ref{crossSI}) dominates and can be estimated as}
\begin{equation}
    \sigma_p^\text{SI} \simeq 1.3\times 10^{-47}~\text{cm}^2~ \left( \dfrac{\theta}{10^{-2}} \right)^2 \left( \dfrac{m_{\tilde \xi}}{500~\text{GeV}} \right)^2 \left( \dfrac{1000~\text{GeV}}{v_R} \right)^2 \,.
    \label{eq:sigmaSI_2}
\end{equation}
{On the other hand, in the limit $\theta \rightarrow 0$, the second term dominates and 
can also be estimated as}
\begin{equation}
    \sigma_p^\text{SI} \simeq 1.8\times 10^{-48}~\text{cm}^2~\left( \dfrac{500~\text{GeV}}{m_{\tilde{\nu}_R}} \right)^4 \left( \dfrac{m_{\tilde \xi}}{500~\text{GeV}} \right)^2\left( \dfrac{1000~\text{GeV}}{v_R} \right)^4\,.
    \label{eq:sigmaSI_1}
\end{equation}
We have verified that the numerical estimates of Eqs.~(\ref{eq:sigmaSI_1}) and~(\ref{eq:sigmaSI_2}) are in agreement within few percents with the public code \texttt{micrOMEGAs}~\cite{Belanger:2018ccd}. 
{Taking the ratio between both equations, the dependence on DM mass disappears and one can straightforwardly deduce that the exotic quark contribution can only be significant for $\theta < 10^{-2}$, $m_{\tilde{\nu}_R}\sim 500$ GeV and $v_R\sim$ 1000 GeV.}

Currently, the most stringent experimental constraints on $ \sigma_p^\text{SI}$ are achieved by the Xenon1T experiment~\cite{Aprile:2018dbl}, which excludes $\sigma_p^\text{SI}\gtrsim5 \times 10^{-46}\,\text{cm}^2$ for a $50$ GeV DM mass and up to  $\sigma_p^\text{SI}\gtrsim 10^{-44}\,\text{cm}^2$ for $10$ TeV DM mass. The sensitivity of the upcoming Darwin experiment~\cite{Aalbers:2016jon} should improve the current bounds from Xenon1T by more than 2 orders of magnitude, and almost will reach the so-called neutrino floor~\cite{Billard:2013qya}. 
{Writing Eq.~(\ref{eq:sigmaSI_2}) as
\begin{equation}
    \sigma_p^\text{SI} \simeq 5.2\times 10^{-47}~\text{cm}^2~ \left( \dfrac{\theta}{10^{-2}} \right)^2 
    \left( \dfrac{m_{\tilde \xi}}{v_R} \right)^2 \,,
    \label{eq:sigmaSI_22}
\end{equation}
one can see that the Xenon1T experiment can already exclude regions of the parameter space with $m_{\tilde \xi} > v_R$ and $\theta > 10^{-2}$.} 
As discussed further on, a sizable part of the remaining viable parameter space should be accessible by the Darwin experiment in the future.

\bigskip

\noindent
{\bf DM-nucleon spin dependent cross section}

\noindent
At the nuclear scale, two kind of effective operators relevant for DM direct detection can be generated from exchange between DM and light quarks of the heavy $Z^\prime$ mediator, as represented in the right diagram of Fig.~\ref{fig:diagram_DD}.\footnote{An additional diagram involving a loop of heavy exotic quarks connected to gluons could also contribute, as for the diagram in the middle of Fig.~\ref{fig:diagram_DD}. However, any relevant gauge invariant effective operator between DM and gluon fields must be of higher order, therefore should be suppressed by the exotic quark masses compared to the right diagram of Fig.~\ref{fig:diagram_DD}. For this reason, this contribution can be safely neglected.} After integrating out the heavy $Z^\prime$ mediator, these spin-dependent (SD) operators are:
\begin{equation}
    \mathcal{O}_{\tilde \xi}^q=C^q_{\tilde \xi} (\bar{\tilde \xi} \gamma_\mu \gamma_5 \tilde \xi)(\bar{q}\gamma^\mu q)\,, \qquad \mathcal{O}_{\tilde \xi}^{q \prime}=C^{q \prime}_\xi (\bar{\tilde \xi} \gamma_\mu \gamma_5 \tilde \xi)(\bar{q}\gamma^\mu \gamma_5 q)\,.
\end{equation}
In the non-relativistic limit, only the operator $\mathcal{O}_{\tilde \xi}^{q \prime}$ gives rise to non-velocity-suppressed contribution. The corresponding Wilson coefficient is given by
\begin{equation}
 C^{q \prime}_{\tilde \xi}\, =\, \dfrac{g_{Z^\prime}^2}{m_{Z^\prime}^2} \dfrac{z(\tilde \xi)}{2} A_q\,,
\end{equation}
where $A_q\equiv(z(q_L)-z(q_R))/2$ is the axial $U(1)^\prime$ charge of a quark $q$. This operator will give rise to the following nucleon-DM effective operator
\begin{equation}
 \mathcal{O}_{\tilde \xi}^{N \prime}\,=\,C^{N \prime}_{\tilde \xi} (\bar{\tilde \xi} \gamma_\mu \gamma_5 \tilde \xi)(\bar{N}\gamma^\mu \gamma_5 N)\,,
\end{equation}
where the corresponding Wilson coefficient can be written as the sum of the spin contribution of the light quarks present within nucleons as
\begin{equation}
    C^{N \prime}_{\tilde \xi} = \sum_{q=u,d,s} C_{\tilde \xi}^{q\prime}  \Delta_q^N = \dfrac{g_{Z^\prime}^2}{m_{Z^\prime}^2} \dfrac{ z(\tilde \xi)}{2} \sum_{q=u,d,s} A_q \Delta_q^N\,,
\end{equation}
with $\Delta_q^N$ the contribution of the quark $q$ to the nucleon $N$ spin. The total nucleon-DM cross section is
\begin{equation}
    \sigma^\text{SD}_N  \, = \,\frac{3 \mu_{\tilde \xi N}^2 z^2(\tilde \xi) g_{Z^\prime}^4}{\pi m_{Z^\prime}^4 }  \left(  \sum_{q=u,d,s} A_q \Delta_q^N \right)^2\, \simeq \, 7.5 \times 10^{-47}~\text{cm}^2 \left( \dfrac{g_{Z^\prime}}{0.1}\right)^4 \left( \dfrac{500~\text{GeV}}{m_{Z^\prime}}\right)^4\,.
\end{equation}

We have checked that this estimate is in agreement within few percents with the public code \texttt{micrOMEGAs}~\cite{Belanger:2018ccd}. {The most stringent bounds on SD interactions are derived by the PICO-60 bubble chamber~\cite{Amole:2017dex,PICO:2019vsc} and Xenon1T experiment~\cite{XENON:2019rxp} which constrain the cross section at the level of $\sigma_{N}^\text{SD} \lesssim 10^{-41}~\text{cm}^2$ for masses $m_{\tilde \xi}\sim 40~\text{GeV}$}. As a result, the above SD cross section is a few of orders of magnitude out of reach of the current generation of experiments, and therefore our U$\mn$ DM scenario remains unconstrained from SD direct searches.

\section{Results}
\label{results}

By using the methods described in previous sections, we evaluate now the current and potential limits on the parameter space of our DM scenario using the relic density constraint, as well as constraints from the LHC and DM direct detection experiments.

\subsection{Scan strategy}

Our DM scenario
is implemented in~\texttt{Feynrules}~\cite{Alloul:2013bka}, exported to \texttt{micrOMEGAs}~\cite{Belanger:2018ccd}, and processed by a private code developed for the numerical analysis performed in~\cite{Lineros:2020eit}. We performed a scan in the parameter space and select the points satisfying the relic density as observed by Planck within a $2 \sigma $ interval around the best fit value, $\Omega_{\tilde \xi} h^2  \in [0.11933 - 2 \times 0.00091, 0.11933 + 2 \times 0.00091]$. The scan was performed by generating random numbers in log-scale {in the ranges of right sneutrino VEV of Eqs.~(\ref{largevev}) and~(\ref{smallvev}), i.e.:
\bea
v_R\in[10, 30 ]~\text{TeV}, \,\,\,\,  v_R\in[1, 10 ]~\text{TeV},
\label{rangess}
\eea
and in the following ranges
of the various parameters:
\begin{align}
& g_{Z^\prime}\, \in \, [g_{Z^\prime}^\text{min},\sqrt{4\pi}] \,, 
&& \theta\,\in \, [10^{-3},10^{-1}] \,, \nonumber \\
& \lambda\, \in \, [\lambda^\text{min},\sqrt{4\pi}]\,,
&& m_{\tilde \nu_R} \, \in \, [0.5, 200]~\text{TeV}\,, \nonumber \\
& k\, \in \, [10^{-3},\sqrt{4\pi}]\,,
&& M'_1 \in \, [0.1, 10]~\text{TeV}  \,, \nonumber\\
& Y_{\mathbb{K}}\, \in \, [Y_{\mathbb{K}}^\text{min},\sqrt{4\pi}] \,.
    \label{scan}
\end{align}
For each of the two intervals in Eq.~(\ref{rangess}) we generate $\sim15$K points satisfying the constraints previously described. 
For $v_R\in[10, 30 ]~\text{TeV}$, the value of $g_{Z^\prime}^\text{min}$ is determined by imposing $m_{Z^\prime}>1~\text{TeV}$, and $m_{Z^\prime}$ is fixed using Eq.~(\ref{entries22}), 
$m_{Z^\prime} = g_{Z^\prime} v_R\sqrt{3}/4$. For $v_R\in[1, 10 ]~\text{TeV}$, taking into account Eq.~(\ref{entries222}) we choose $g_{Z^\prime}^\text{min}=10^{-3}$ and the $Z^\prime$ mass is scanned over in the range 
$m_{Z^\prime}\in [\text{max}(1\,\text{TeV}, g_{Z^\prime} v_R\sqrt{3}/4), 20\,\text{TeV}]$.
The upper bound on $g_{Z^\prime}$ (and on the other dimensionless couplings $\lambda$, $k$, $Y_{\mathbb{K}}$) is chosen as conservative as possible, just imposing perturbativity at the electroweak scale.

\begin{figure}[t!]
    \centering
    \includegraphics[width=0.48\linewidth]{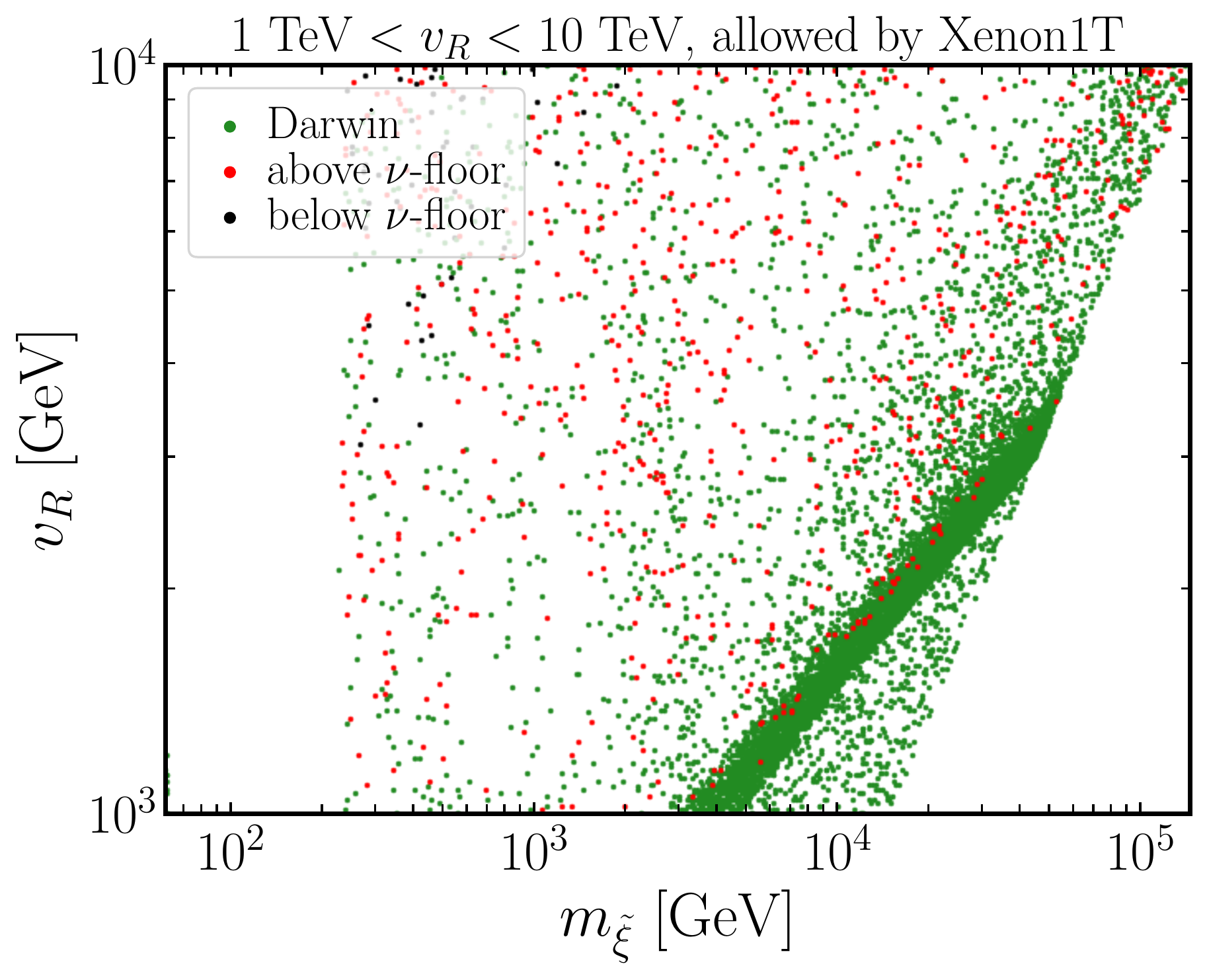}
    \includegraphics[width=0.48\linewidth]{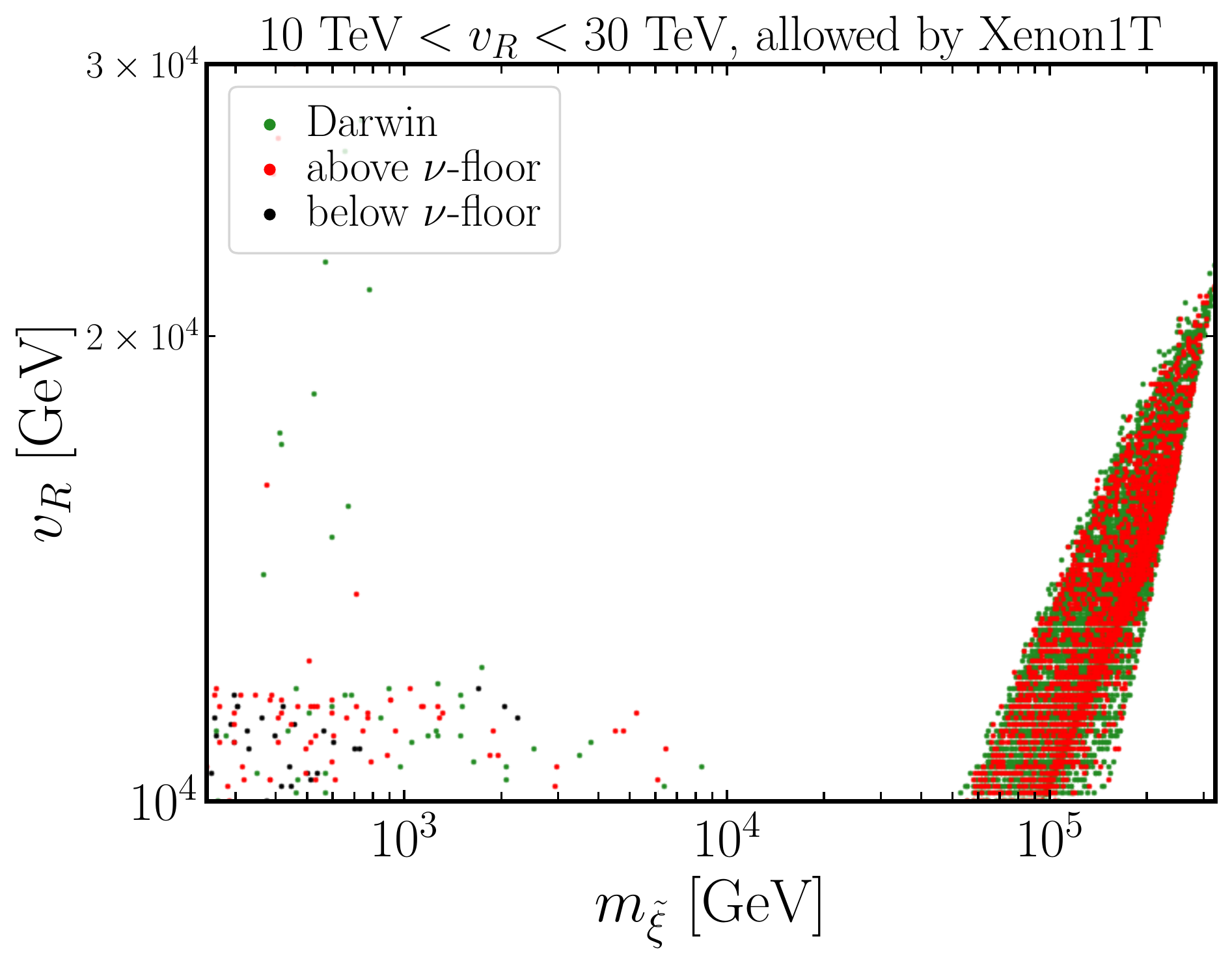}
    \caption{
    Allowed values of the right sneutrino VEV $v_R$ in the parameter space of our model versus DM mass $m_{\tilde\xi}$, for the relevant two intervals (left panel) $1~\text{TeV}<v_R<10~\text{TeV}$, and (right panel) $v_R>10$ TeV. 
    All points represented correspond to a direct detection cross section compatible with constraints from Xenon1T experiment~\cite{Aprile:2018dbl}.
    Green dots will be probed by the upcoming Darwin
    experiment~\cite{Aalbers:2016jon}. 
    Red (black) dots correspond to points above (below) the neutrino 
    floor~\cite{Billard:2013qya} (see also Fig.~\ref{fig:scan_direct_detection}).}
    \label{fig:scan_parameter_space_VR_VS_MDM}
\end{figure}

\begin{figure}[t!]
    \centering
    \includegraphics[width=0.48\linewidth]{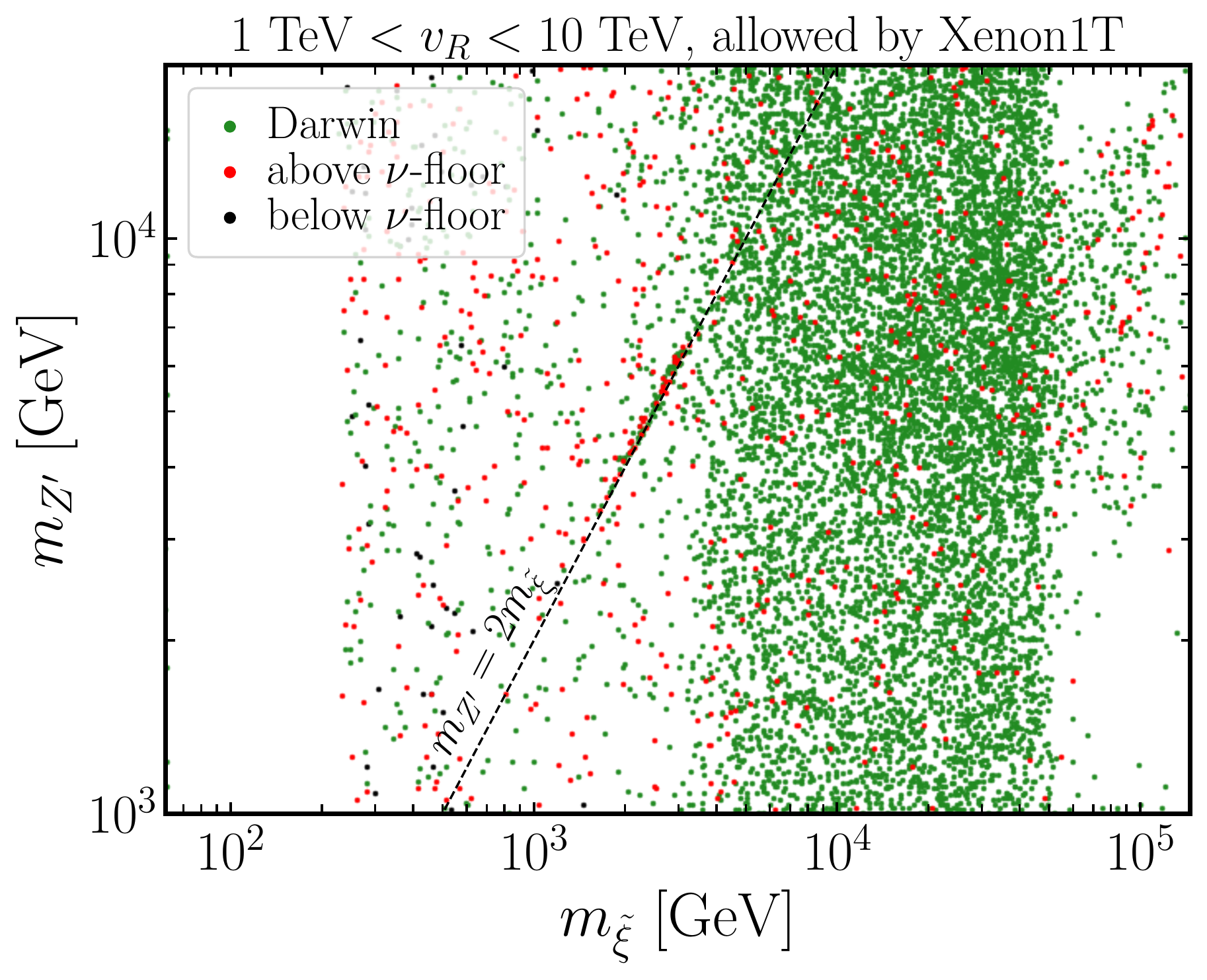}
    \includegraphics[width=0.48\linewidth]{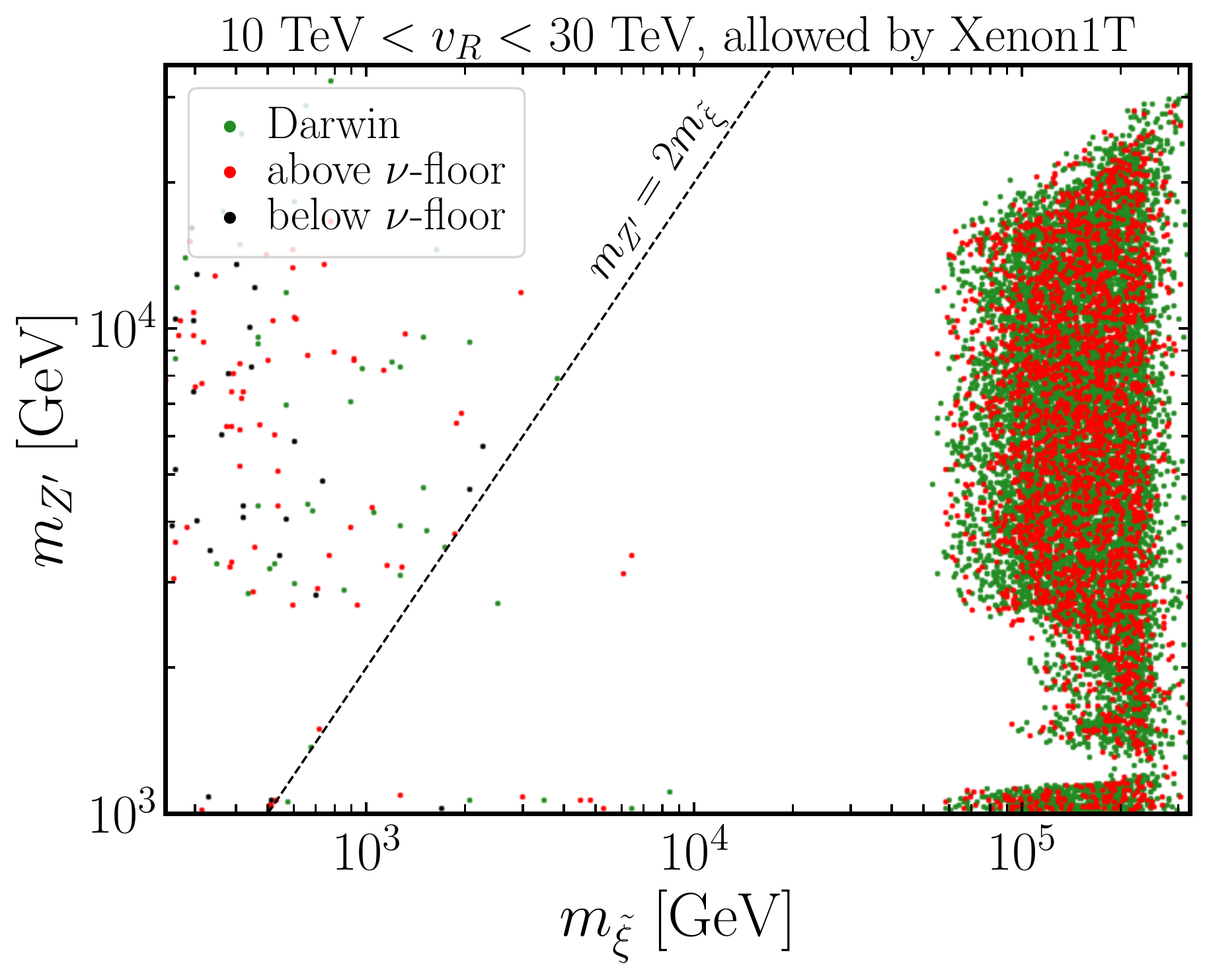}
    \caption{The same as in Fig.~\ref{fig:scan_parameter_space_VR_VS_MDM}, but showing the mass of the $U(1)'$ gauge boson $m_{Z'}$ versus DM mass $m_{\tilde\xi}$.}
    \label{fig:scan_parameter_space_MZP_VS_MDM}
\end{figure}

From the randomly chosen values of the above parameters,
we can deduce the masses of the heavy new particles, 
DM and exotic quarks, using the second and third formulas of Eq.~(\ref{masses2}).
In addition, we fix neutralino masses by setting one of the three RH neutrino masses to a negligible value as discused in Sec.~\ref{scenarios}, the second one is fixed 
using Eq.~(\ref{massesneu2}), and the remaining one as well as the $\tilde Z'$ mass are fixed according to Eq.~(\ref{eigen}). The charged and neutral Higgsino masses are set to the value of the $\mu$-term (see
Eqs.~(\ref{massesneuchar1}) and~(\ref{massesneuchar})) as expressed in the first formula of Eq.~(\ref{masses2}). 
In Eq.~(\ref{scan}), the value of the coupling $Y_{\mathbb{K}}^\text{min}$ is determined by imposing $m_{\mathbb{K}}>1200\text{ GeV}$ from LHC searches of R-hadrons, as discussed in Sec.~\ref{constraintslhc}. The value of the coupling $\lambda^\text{min}$ is determined by imposing $\mu>100 \text{ GeV}$ in order to fulfill the chargino bound. In addition, we assume for simplicity that the masses of the three right sneutrinos are the same.

\begin{figure}[t!]
    \centering
    \includegraphics[width=0.48\linewidth]{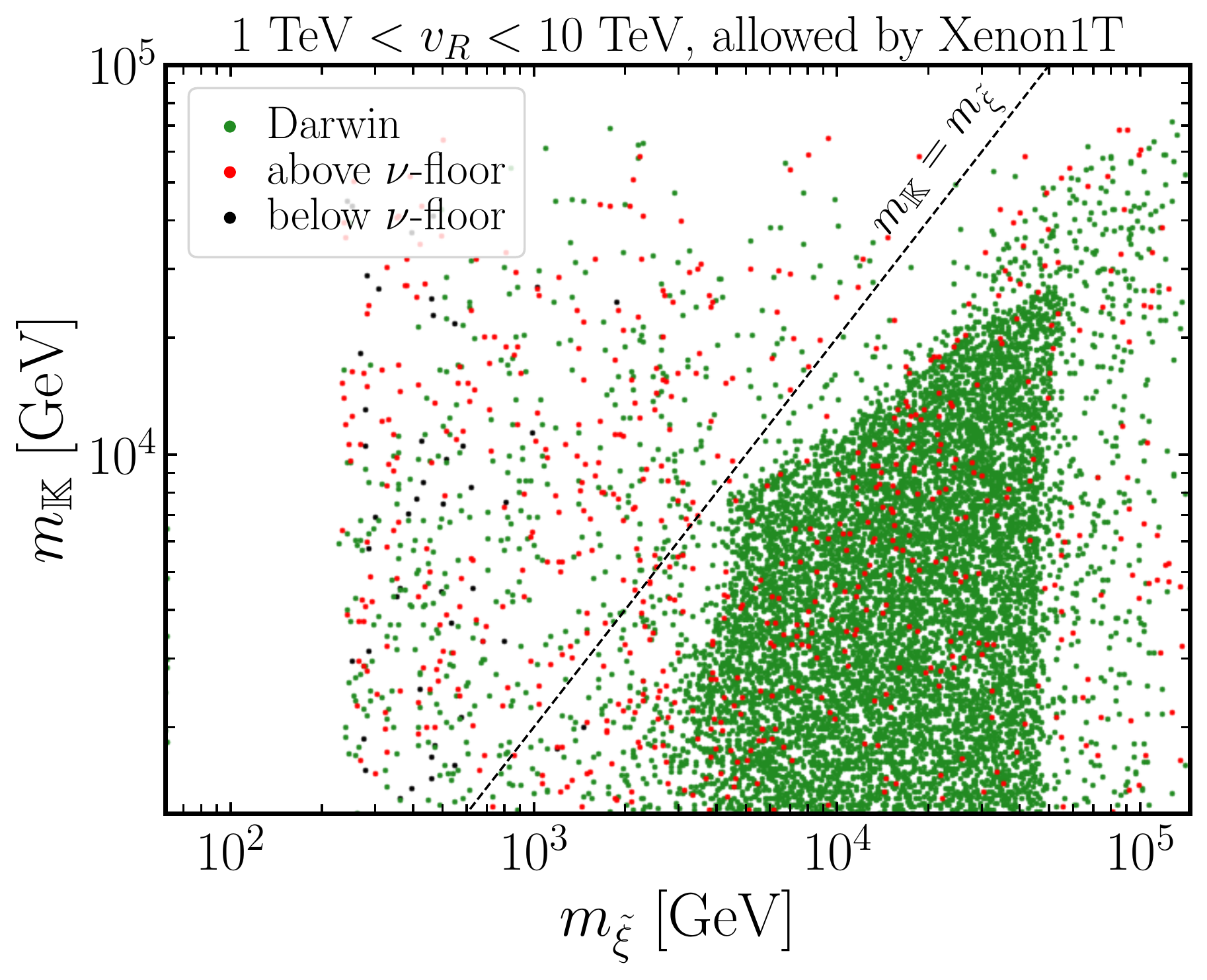}    
    \includegraphics[width=0.48\linewidth]{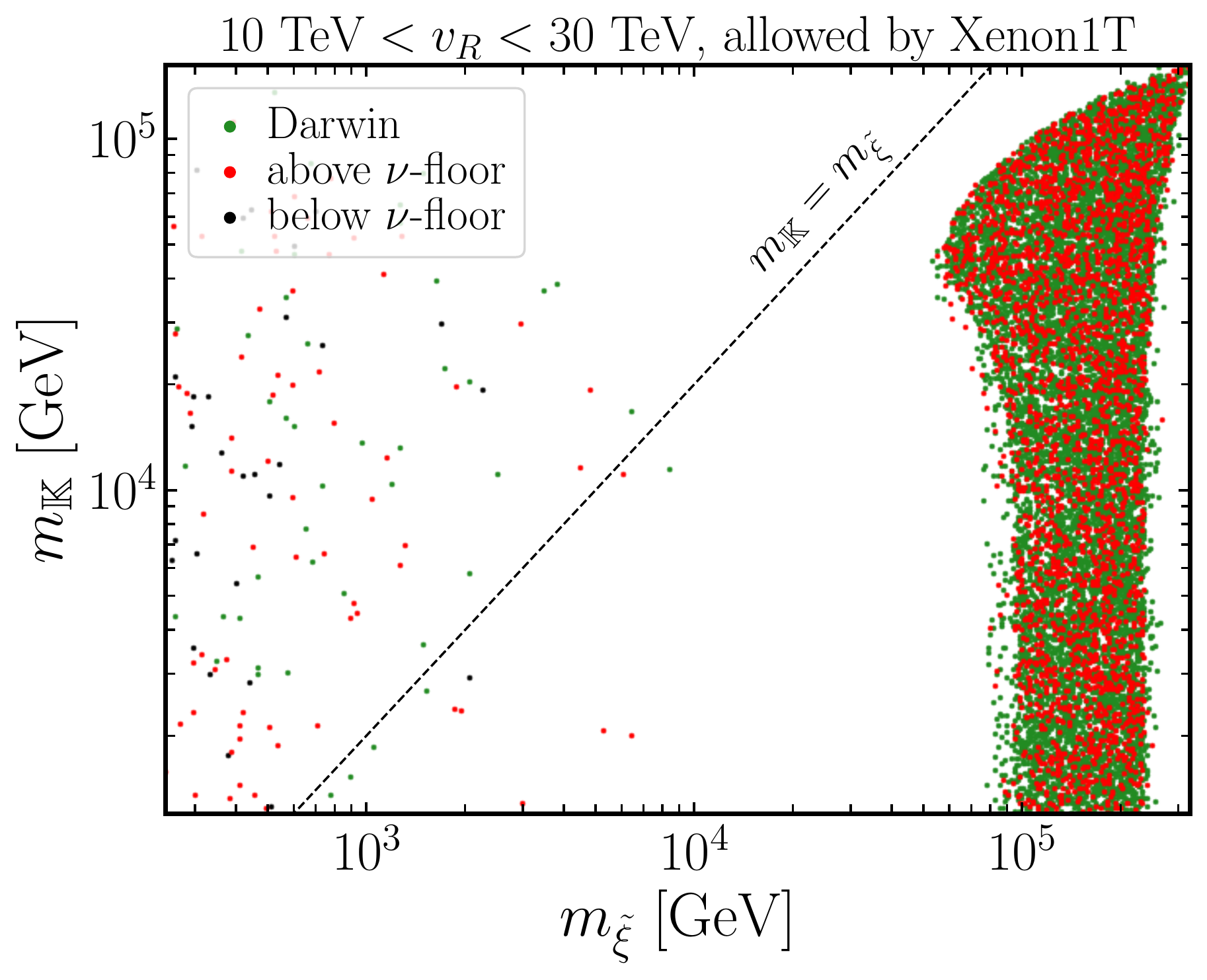}
    \caption{The same as in Fig.~\ref{fig:scan_parameter_space_VR_VS_MDM}, but showing the mass of the exotic quarks $m_{\mathbb{K}}$ versus DM mass $m_{\tilde\xi}$.}
    \label{fig:scan_parameter_space_MK_VS_MDM}
\end{figure}

\begin{figure}[t!]
    \centering
    \includegraphics[width=0.48\linewidth]{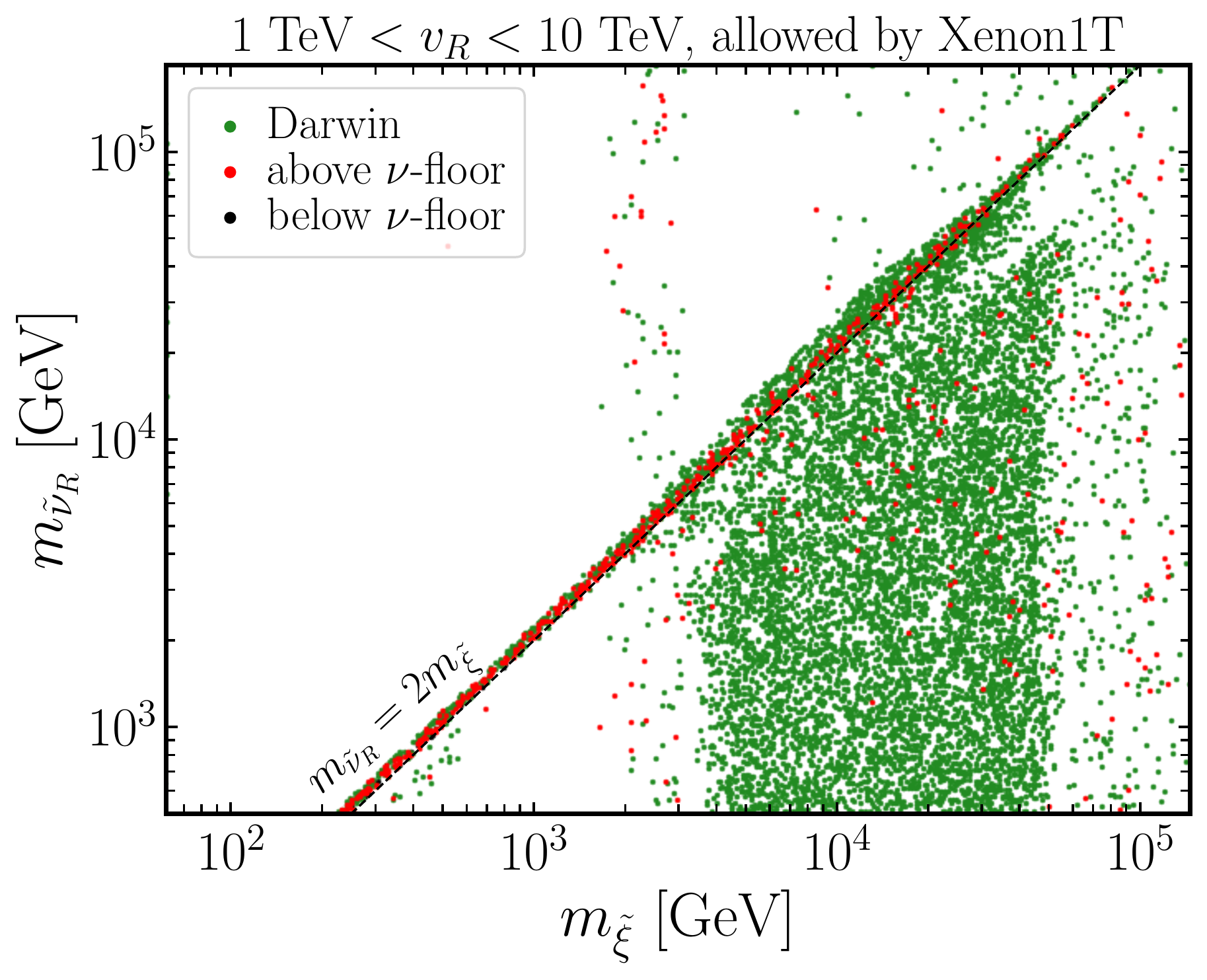}    
    \includegraphics[width=0.48\linewidth]{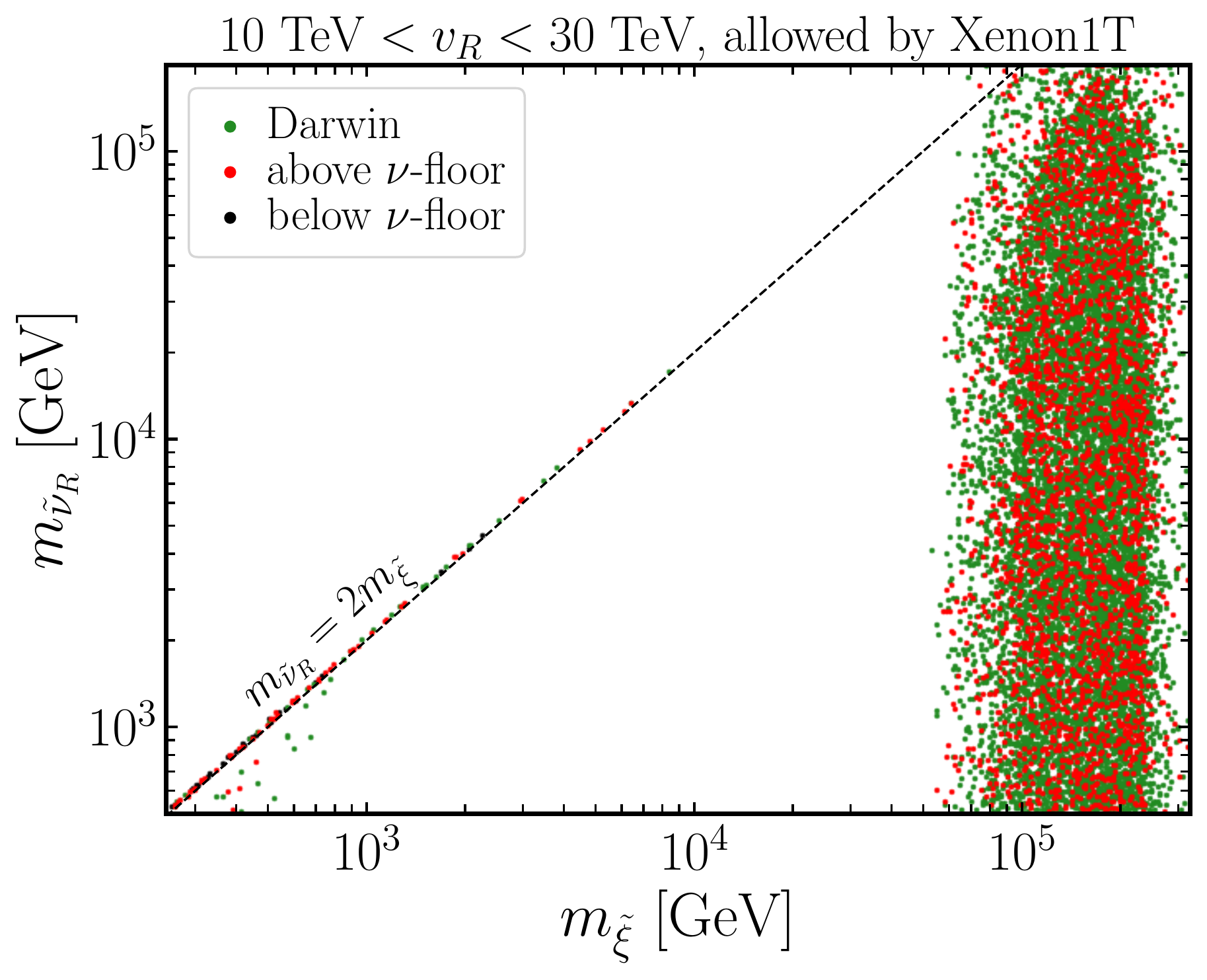}
    \caption{The same as in Fig.~\ref{fig:scan_parameter_space_VR_VS_MDM}, but showing the mass of the right sneutrinos $m_{\tilde\nu_R}$ versus DM mass $m_{\tilde\xi}$.}
    \label{fig:scan_parameter_space_MSNUR_VS_MDM}
\end{figure}

\subsection{Numerical analysis}
\label{numerical}

\paragraph{$\mathbf{10~\text{TeV} < v_R < 30~\text{TeV}}$.}
These high values of the VEV $v_R$ tend to push the viable corner of the parameter space close to the perturbative unitarity limit for the DM annihilation cross section.
As some of the most relevant new states ($\tilde \xi, \mathbb{K}, Z^\prime$) acquire their masses from the VEV of the right sneutrinos, some contributions to the cross section schematically behave as $\langle \sigma v_{\tilde \xi} \rangle \propto v_R^{-2}$ (such as in Eqs.~(\ref{eq:sigmav_KK_simp}),~(\ref{eq:sigmavscalars}) and~(\ref{eq:sigmavZpnur})), which require a coupling that has to be larger for larger values of the VEV $v_R$, in order to achieve the correct relic abundance.
Indeed, for this part of the parameter space the typical values for the VEV remain mostly below $v_R \lesssim 20$ TeV, as illustrated in the right panel of Fig.~\ref{fig:scan_parameter_space_VR_VS_MDM}. As the $Z^\prime$ mass cannot be much larger than $v_R$ (see Eq.~(\ref{entries22})), the values of $m_{Z^\prime}$ are typically $m_{Z^\prime}\lesssim 10-20$ TeV as shown in the right panel of Fig.~\ref{fig:scan_parameter_space_MZP_VS_MDM}. 
(We also show in Figs.~\ref{fig:scan_parameter_space_MK_VS_MDM} and~\ref{fig:scan_parameter_space_MSNUR_VS_MDM}
the exotic quark and right sneutrino masses versus DM mass, to be discussed below.)
Since the range of $v_R$ spanned by the scan is rather narrow, there is almost a one-to-one relation between DM mass and the coupling $k$ (see the second formula of Eq.~(\ref{masses2})), which can be seen as a straight line in the right panel of Fig.~\ref{fig:scan_dimensionless_couplings}, where one can also distinguish clearly 2 regimes, at low and high DM masses.

\begin{figure}[t!]
    \centering
    \includegraphics[width=0.48\linewidth]{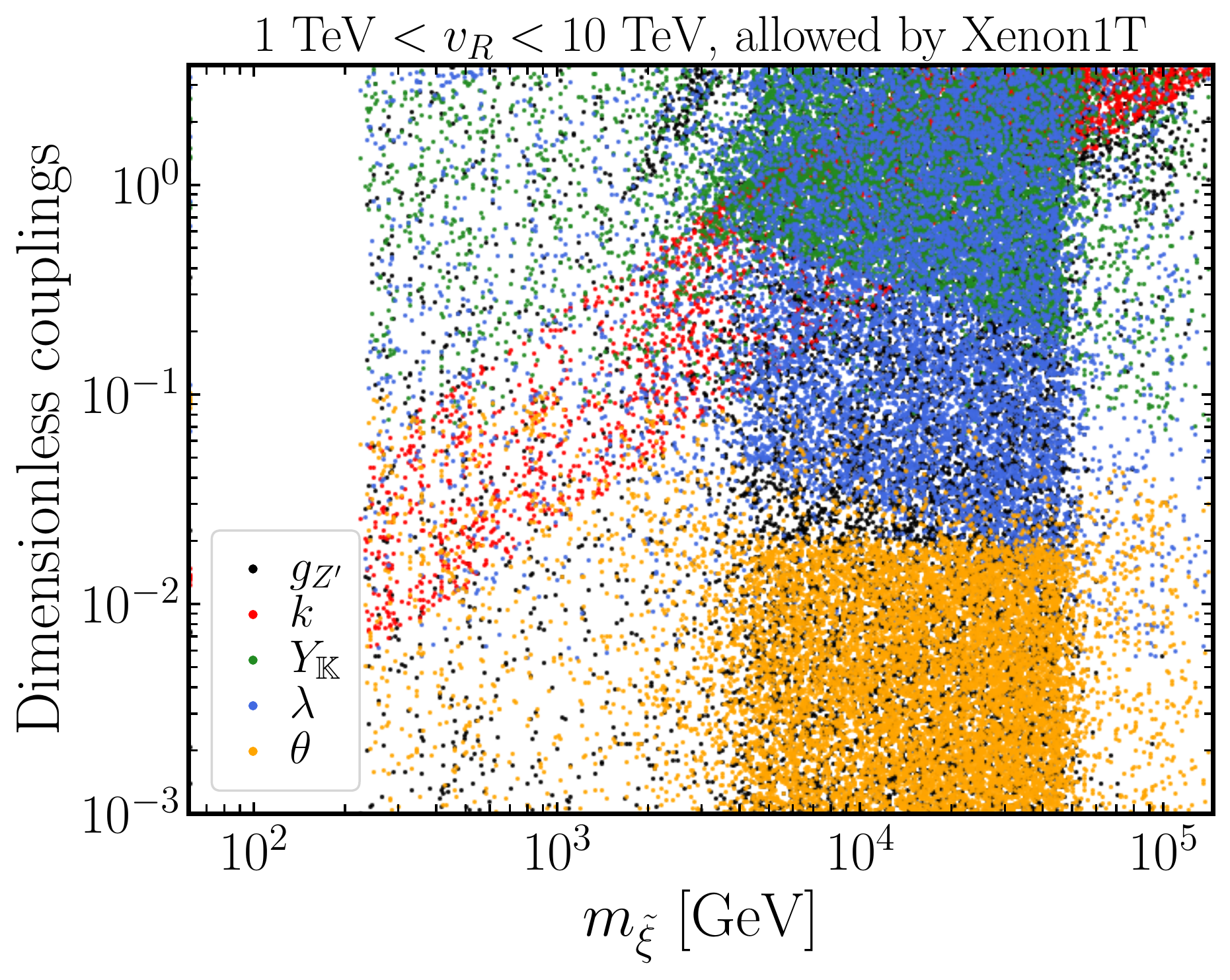}    
    \includegraphics[width=0.48\linewidth]{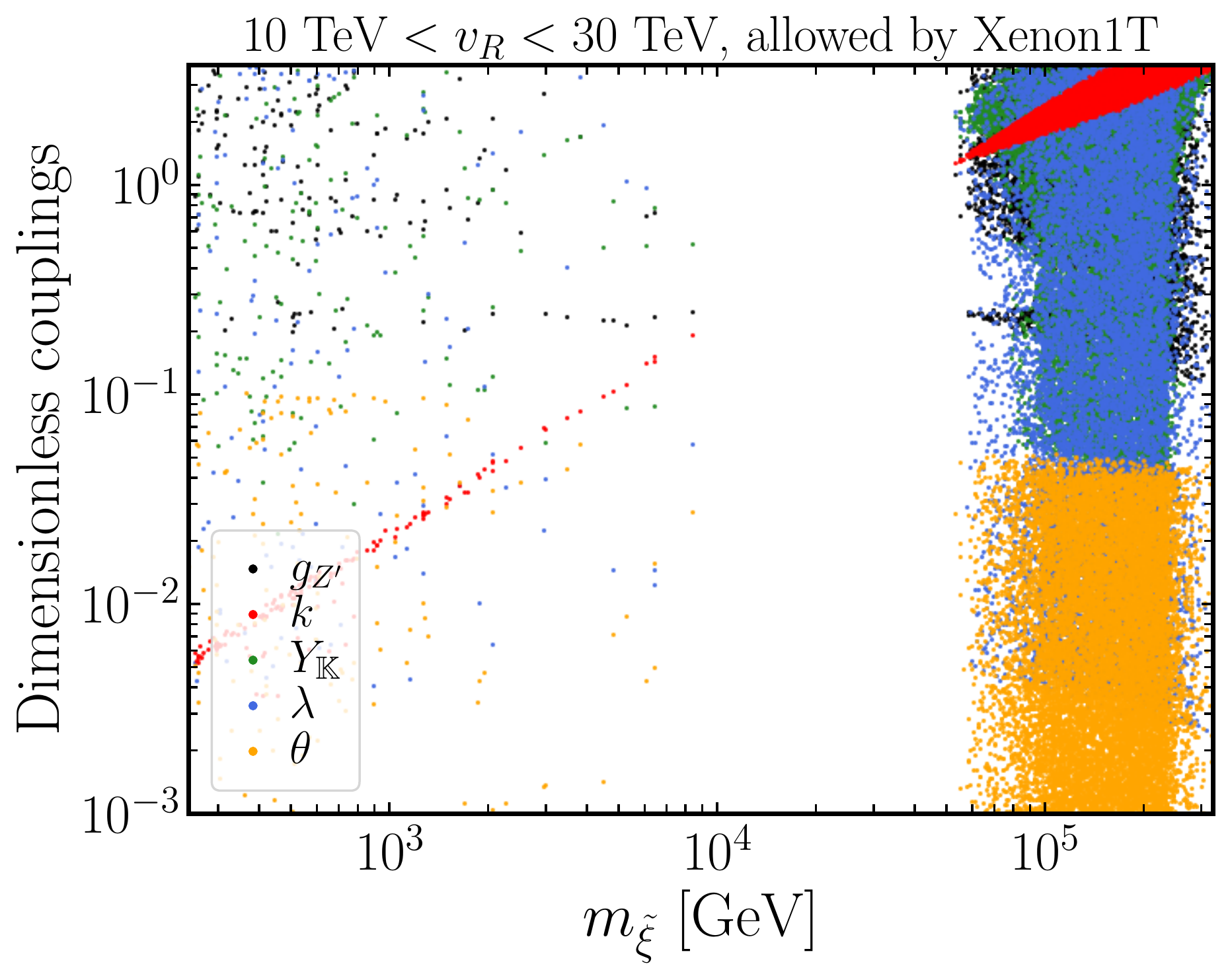}
    \caption{
Values of the dimensionless couplings in the parameter space of our model versus DM mass $m_{\tilde\xi}$, for the relevant two intervals of right sneutrino VEVs (left panel) $1~\text{TeV}<v_R<10~\text{TeV}$, and (right panel) $v_R>10$ TeV. All points represented correspond to a direct detection cross section compatible with constraints from Xenon1T experiment~\cite{Aprile:2018dbl}.}
    \label{fig:scan_dimensionless_couplings}
\end{figure}

\begin{figure}[t!]
    \centering
    \includegraphics[width=0.48\linewidth]{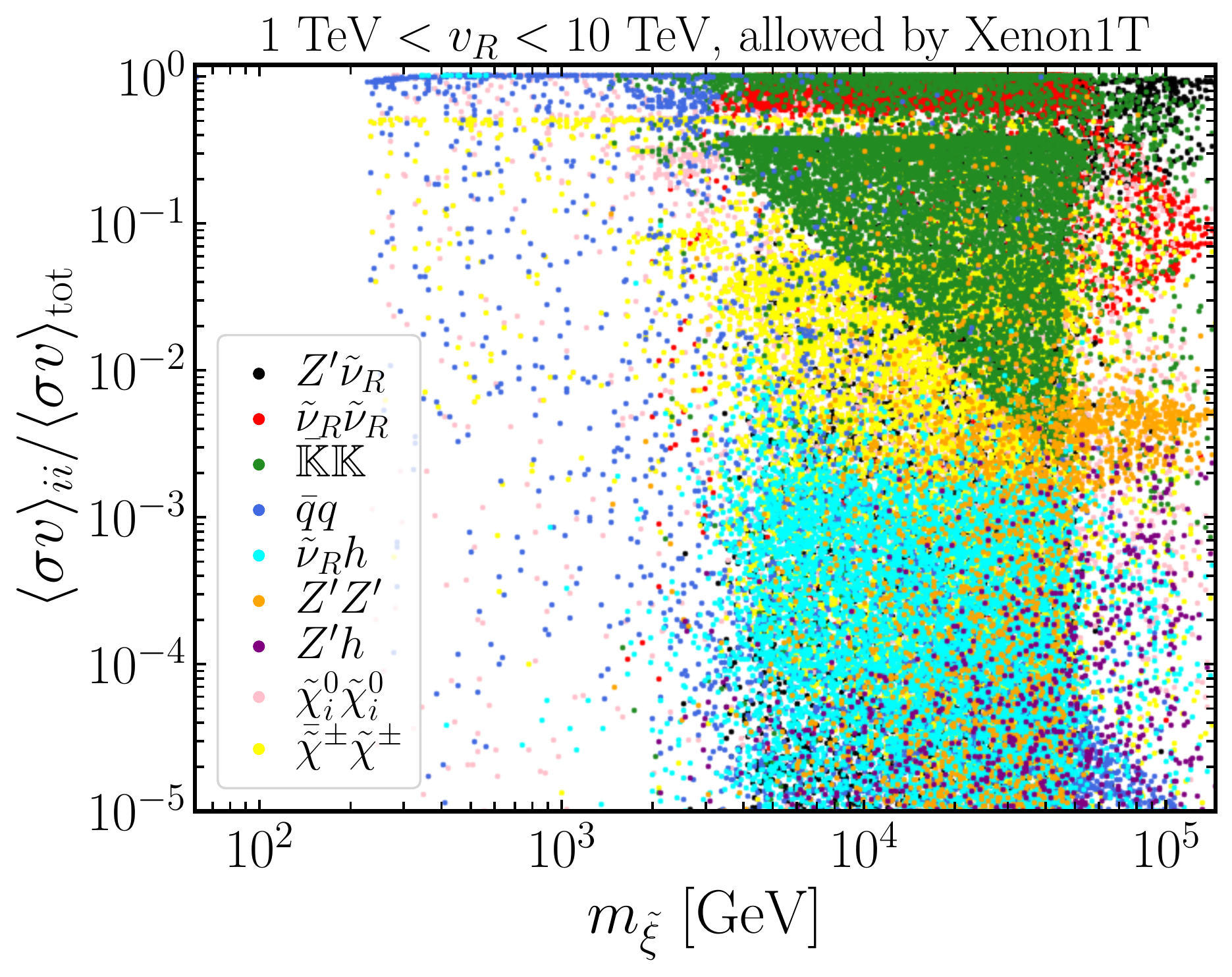}
    \includegraphics[width=0.48\linewidth]{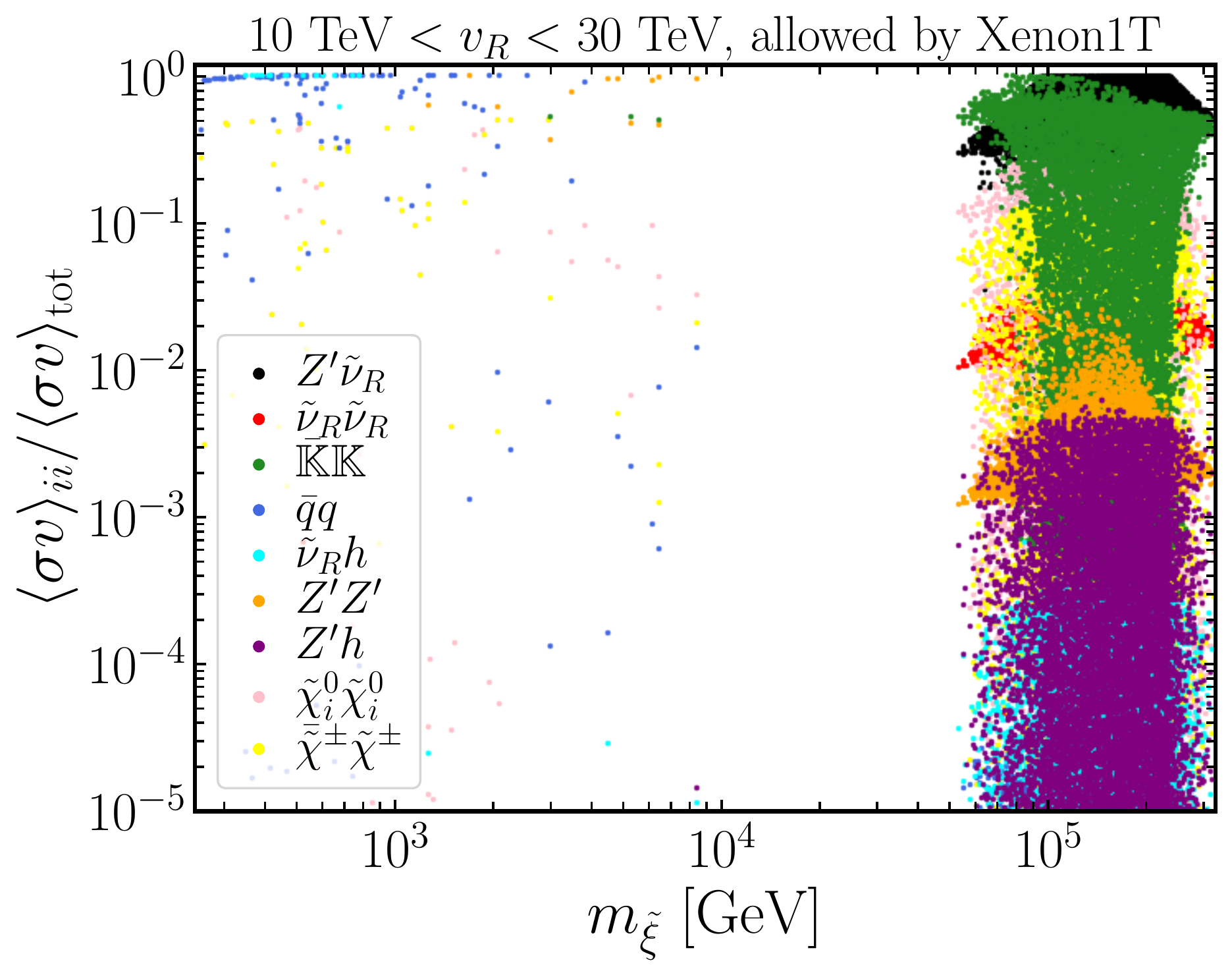}
    \caption{
   The same as in  Fig.~\ref{fig:scan_dimensionless_couplings}, but showing the
    relative contribution of each annihilation channel to the overall cross section
    versus DM mass $m_{\tilde\xi}$.
    }
    \label{fig:scan_annihilation_channels}
\end{figure}

\begin{figure}[t!]
    \centering
    \includegraphics[width=0.48\linewidth]{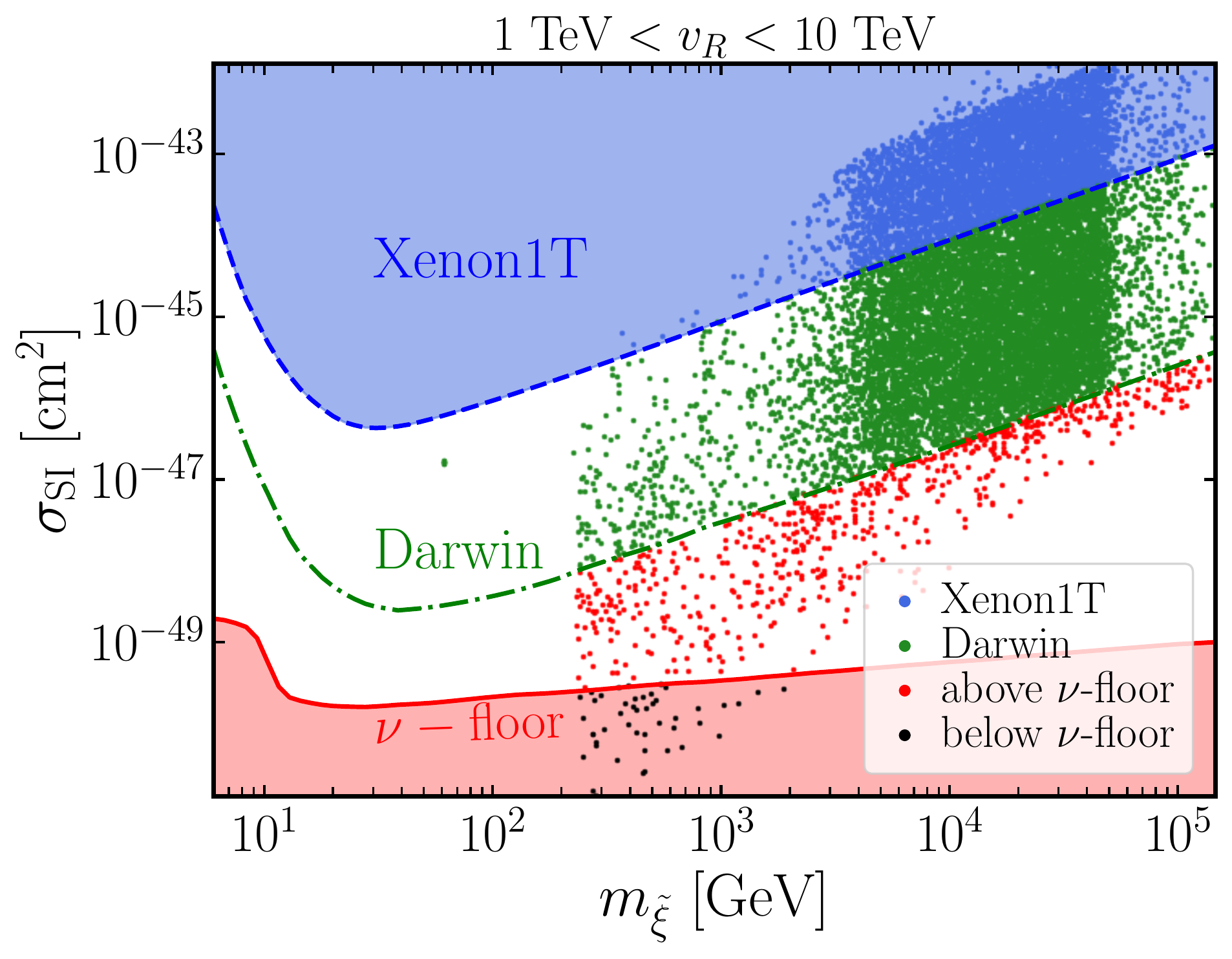}    
    \includegraphics[width=0.48\linewidth]{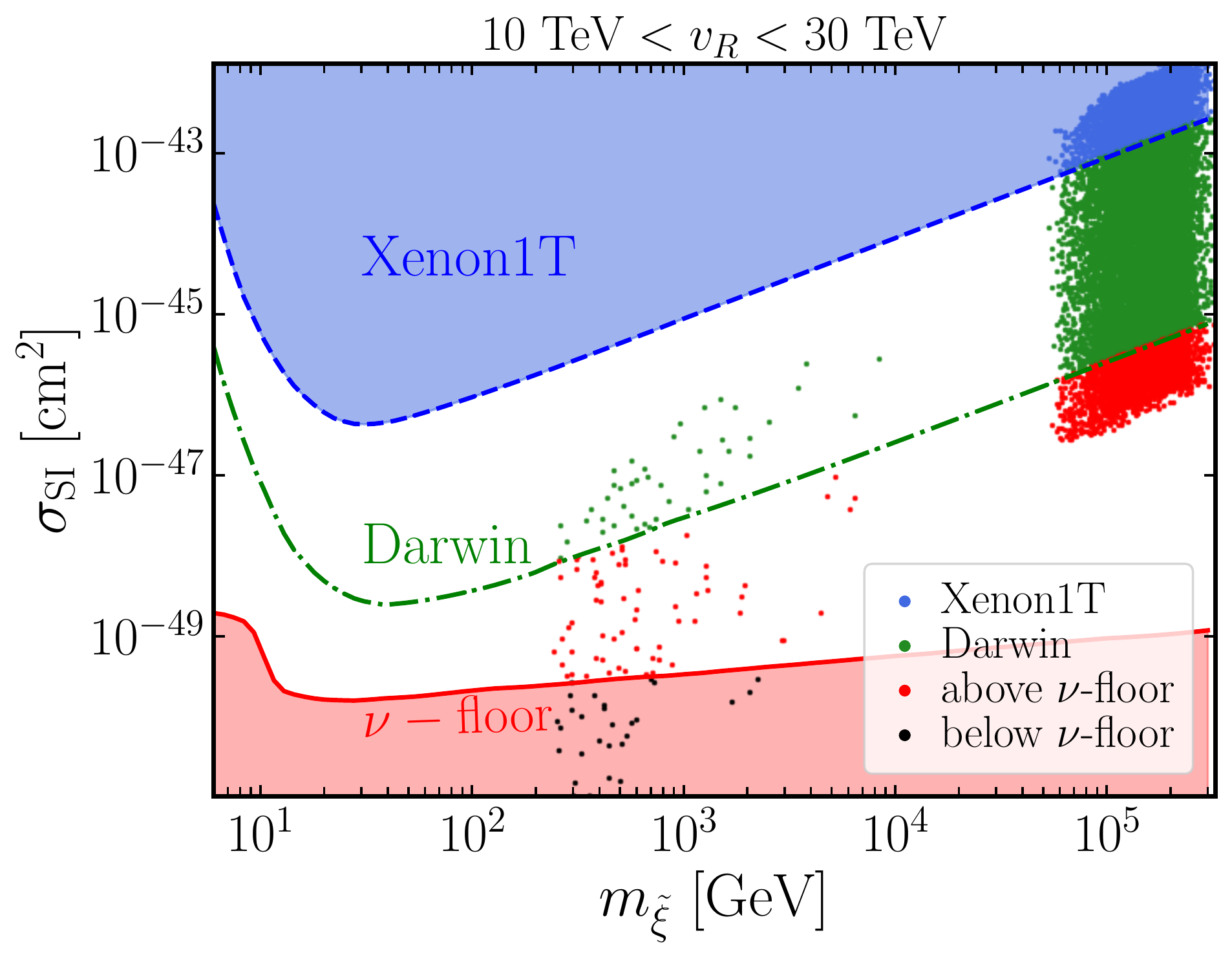}
    \caption{
    Spin-independent DM-nucleon cross section $\sigma_\text{SI}$ versus DM mass $m_{\tilde\xi}$ in the parameter space of our model for the relevant two intervals of right sneutrino VEVs (left panel) $1~\text{TeV}<v_R<1~\text{TeV}$, and (right panel) $v_R>10$ TeV.
    Blue dots are excluded by the Xenon1T experiment~\cite{Aprile:2018dbl}.
    Green dots will be probed by the upcoming Darwin
    experiment~\cite{Aalbers:2016jon}. 
    Red (black) dots correspond to points above (below) the neutrino 
    floor~\cite{Billard:2013qya}.}
    \label{fig:scan_direct_detection}
\end{figure}

\begin{figure}[t!]
    \centering
    \includegraphics[width=0.48\linewidth]{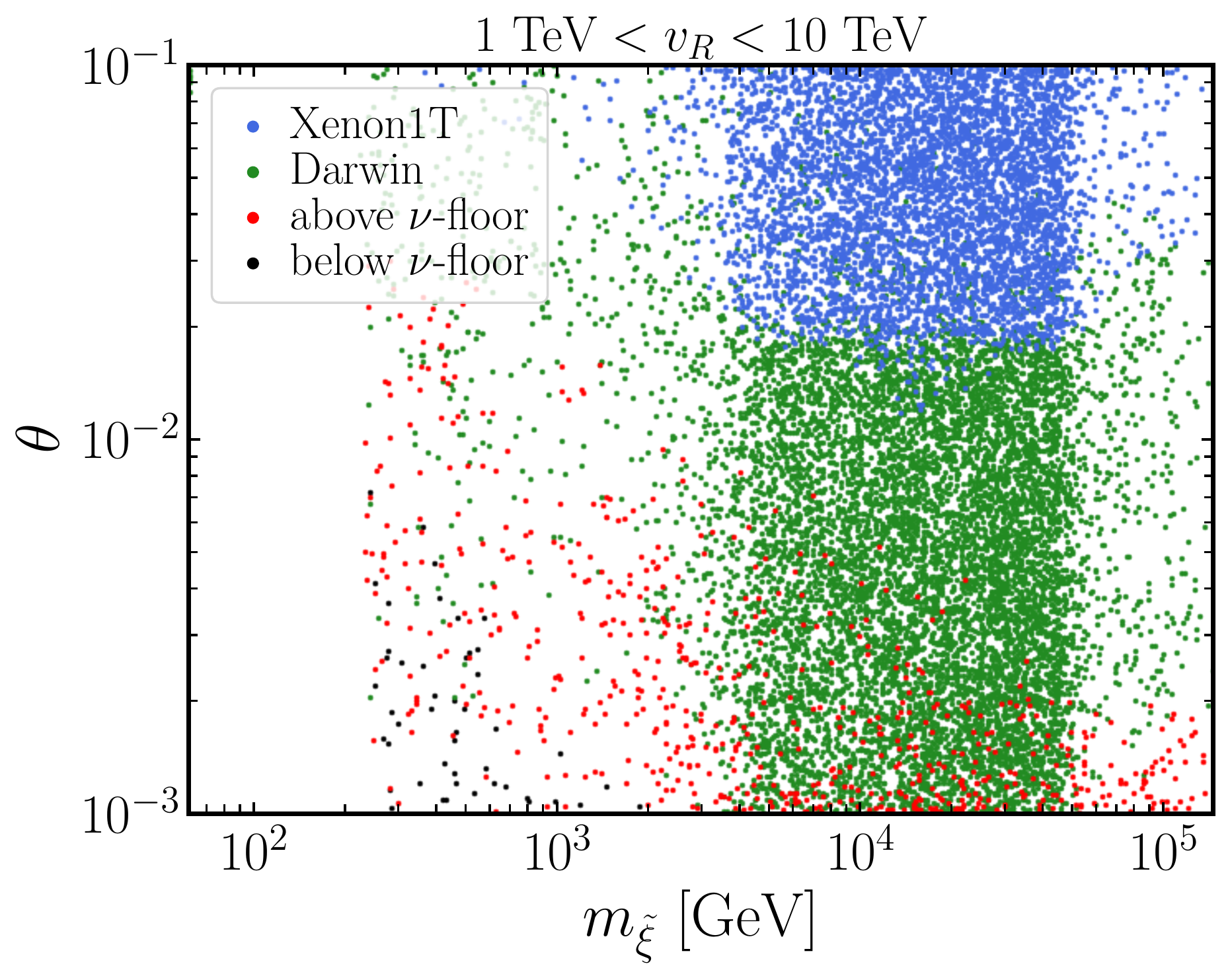}    
    \includegraphics[width=0.48\linewidth]{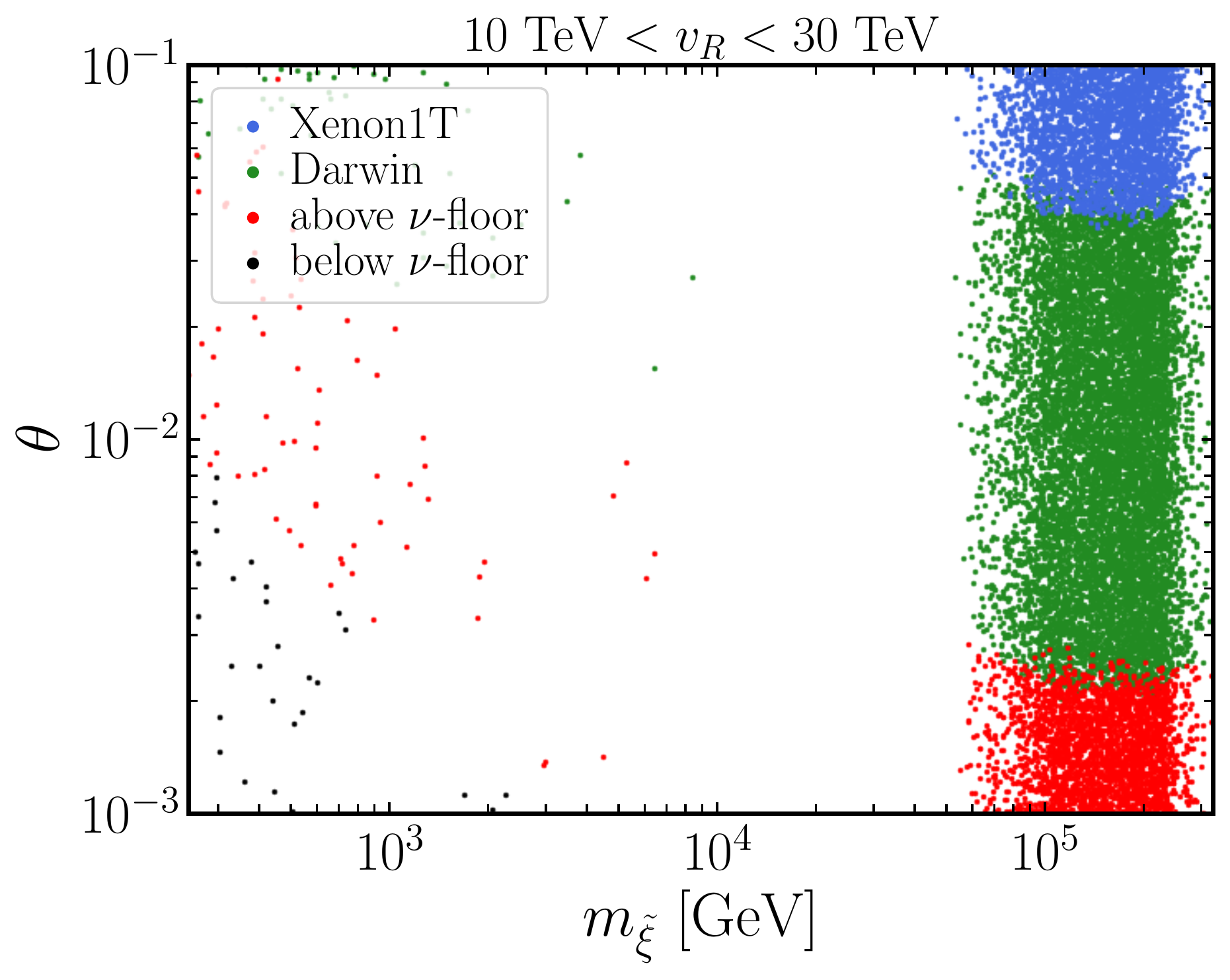}
    \caption{
    The same as in Fig.~\ref{fig:scan_direct_detection}, but showing the mixing angle $\theta$ between the right sneutrinos and the SM-like Higgs
    versus DM mass $m_{\tilde\xi}$.}
    \label{fig:scan_direct_theta}
\end{figure}

For large DM masses $m_{\tilde \xi} \gtrsim 10^5 $ GeV, the coupling $k$ becomes sufficiently large so as to allow predominant DM annihilation to $\tilde \xi \tilde \xi \rightarrow Z' \tilde \nu_R$.
In addition, annihilation to exotic quarks $\bar{\mathbb{K}} \mathbb{K}$ becomes kinematically allowed since $m_{\tilde \xi}>m_\mathbb{K}$ as shown in the right panel of Fig.~\ref{fig:scan_parameter_space_MK_VS_MDM}, and typically contributes significantly to the total cross section.
{The importance of these channels with respect to the rest can be seen in the right panel of Fig.~\ref{fig:scan_annihilation_channels}.}
This regime offers interesting detection prospects. As illustrated in the right panel of Fig.~\ref{fig:scan_direct_detection}, the Xenon1T experiment~\cite{Aprile:2018dbl} excludes already a subdominant part of the parameter space.
{This region corresponds to DM mass larger 
than $v_R$ and $\theta > 10^{-2}$ (see Fig.~\ref{fig:scan_direct_theta}) in agreement with our discussion of Eq.~(\ref{eq:sigmaSI_22})}
while a larger proportion should be in the reach of the upcoming Darwin experiment~\cite{Aalbers:2016jon}. In addition, in this regime the main DM annihilation channels are not velocity suppressed, as can be seen respectively in Eqs.~(\ref{eq:sigmavZpnur}) and~(\ref{eq:sigmav_KK_simp}), which could allow for indirect detection signals. Indeed, as the cross section is velocity independent, frequent DM annihilations within the galactic halo could produce a large flux of $Z^\prime$ bosons that will subsequently decay into SM quarks and generate potentially large gamma-ray and antiproton signals. The strongest bounds on DM annihilations are typically achieved for the $\bar b b$ final state, and are derived from combined searches towards dwarf spheroidal galaxies with Fermi-LAT, HAWC, H.E.S.S., MAGIC, and VERITAS, yielding $m_{\tilde \xi}\gtrsim 100~\text{GeV}$~
\cite{Hess:2021cdp}. Recent analyses of AMS-02 antiproton data~\cite{Kahlhoefer:2021sha,Calore:2022stf} have shown that AMS-02 can constrain DM annihilations to $\bar b b$ with the bound $m_{\tilde \xi}\gtrsim \mathcal{O}(50)~\text{GeV}$ but also at higher masses around $m_{\tilde \xi}\sim \mathcal{O}(300)~\text{GeV}$. Therefore, we expect these constraints to affect very marginally our parameter space, as only very few points can reproduce the correct relic abundance for $m_{\tilde \xi} \lesssim 400~\text{GeV}$. The upcoming ground based CTA telescope network should reach a sensitivity to DM annihilation of $\mathcal{O}(10^{-27}-10^{-26})\, \text{cm}^3\text{s}^{-1}$ for DM masses $m_{\tilde \xi}\sim \mathcal{O}(1-10)\, \text{TeV}$ and $\mathcal{O}(10^{-26}-10^{-25})\, \text{cm}^3\text{s}^{-1}$ for $m_{\tilde \xi}\sim \mathcal{O}(10^2) \, \text{TeV}$ depending on the 
specific annihilation channel~\cite{CTAConsortium:2018tzg,Pierre:2014tra,Silverwood:2014yza,Lefranc:2015pza}. As a result, we might expect CTA to probe a part of the parameter space of this regime and would most likely be accessible by a telescope with an improved sensitivity. The predicted indirect and direct detection signals could allow to identify the DM candidate and discriminate our scenario from other models. {On more theoretical grounds, such a regime requires at least one of the couplings, typically $k$, to be rather large $\mathcal{O}(1)$ or even larger as illustrated in the right panel of Fig. \ref{fig:scan_dimensionless_couplings}, pushing the model towards the limits of validity for perturbative unitarity for the highest DM masses.}

For smaller values of the DM mass $m_{\tilde \xi}<10$ TeV, the coupling $k$ is smaller and the only possibility to achieve the correct relic density is to rely on $s-$channel resonant annihilation induced by a $\tilde \nu_R$-mediator, as can be seen in the right panel of Fig.~\ref{fig:scan_parameter_space_MSNUR_VS_MDM}. 
{Note in particular that for $ m_{\tilde \xi}> 1$ TeV, as the cross section is typically even more suppressed, the resonance $m_{\tilde \nu_R }\simeq 2 m_{\tilde \xi}$ must be narrower to allow the cross section to reach the value reproducing the correct relic density. This implies that most of the couplings must be relatively small simultaneously to satisfy this condition, which is less frequently achieved in the scan up to a point where the resonance is no longer enough to get a right value of the annihilation cross section. This explains the apparent mass gap for $ m_{\tilde \xi}\sim 10-100$ TeV and the rarefaction of the points for increasing $m_{\tilde \xi}$ along the $m_{\tilde \nu_R }= 2 m_{\tilde \xi}$ line in the right panel of Fig.~\ref{fig:scan_parameter_space_MSNUR_VS_MDM}.}
From the right panel of Fig.~\ref{fig:scan_annihilation_channels}, one can also see that the most efficient annihilation channels for this regime are $\bar q q$,  $\bar{\mathbb{K}} \mathbb{K}$, $Z^\prime Z^\prime$ and  $\tilde \nu_R h$ depending on the specific values of the various dimensionless parameters. The DM annihilation cross sections are not velocity suppressed only for $\bar{\mathbb{K}} \mathbb{K}$ and $Z^\prime Z^\prime$, as shown in Eqs.~(\ref{eq:sigmav_KK_simp}) and~(\ref{eq:sigmav_ZpZp}), respectively, therefore potentially CTA could probe some part of the parameter space, provided that the produced gamma-ray spectrum should be similar\footnote{A dedicated analysis of the expected gamma-ray spectrum would be required to make a stronger statement, which goes beyond the scope of this paper.} to that of $W^+W^-$ or $\bar b b$~\cite{CTAConsortium:2018tzg,Pierre:2014tra,Silverwood:2014yza,Lefranc:2015pza}. Direct searches offer less interesting detection prospects for this regime as they rely on couplings which can be smaller than in the heavy DM regime. Indeed, some part of the parameter space corresponds to a SI cross section out of reach of the future Darwin experiment and a subdominant portion below the neutrino floor~\cite{Billard:2013qya}, as shown in Fig.~\ref{fig:scan_direct_detection}. Nevertheless, Darwin should still be able to probe a non negligible part of the parameter space for this regime. {Moreover, the smaller couplings predicted for this regime allow to relax possible tensions with perturbativity that appear for very large DM masses. In particular, for this corner of the parameter space the coupling $k$ is always smaller or much smaller than one, as illustrated in the right panel of Fig. \ref{fig:scan_dimensionless_couplings}.}

\par \medskip

\paragraph{$\mathbf{1~\text{TeV} < v_R < 10~\text{TeV}}$.} Similarly to the previous case, for this interval one can also distinguish two main regimes: for DM masses above and below $m_{\tilde \xi}\sim 3$ TeV.
For large DM masses we recover a similar pattern for the coupling $k$, which typically grows with the DM mass as can be seen in the left panel of Fig.~\ref{fig:scan_dimensionless_couplings}. However, since the VEV interval covers one order of magnitude, the points are more scattered than for the case $v_R>10$ TeV. For this regime, the DM is heavier than the $Z^\prime$ and the relevant dominant annihilation channels are $\tilde \xi \tilde \xi \rightarrow \tilde \nu_R \tilde \nu_R, Z^\prime \tilde \nu_R$ as well as $\bar{\mathbb{K}} \mathbb{K}$ in the final state. Since only the $\tilde \nu_R \tilde \nu_R$ annihilation channel is velocity suppressed, most of this regime should be in the reach of indirect searches with the upcoming CTA as discussed previously, and as shown in several analyses~\cite{CTAConsortium:2018tzg,Pierre:2014tra,Silverwood:2014yza,Lefranc:2015pza}. As for the case with $v_R>10$ TeV and large DM masses, the Darwin experiment will also play a determinant role in constraining or discovering DM in this regime for 
$m_{\tilde \xi}\gsim 3$ TeV. This is shown in the left panel of
Fig.~\ref{fig:scan_direct_detection}.
Let us remark that only a small bunch of points (1\% of the total) corresponds to the case of dominance of the exotic quark contribution to the cross section (\ref{crossSI}). These points are in the Darwin region with DM mass in the range between about 3 TeV and 40 TeV, and their corresponding parameters are close to the lower limit of the scan, i.e. $\theta\sim 10^{-3}$, $m_{\tilde{\nu}_R}\sim 500$ GeV and $v_R\sim$ 1000 GeV.

{Interestingly, even if this contribution is typically not dominant for our choice of scan range, such contribution would become relevant for values of $\theta$ smaller than the lower limit of our scan ($10^{-3}$) and offers a possibility of probing this part of the parameter space in the future. This is illustrated in the left panel of Fig.~\ref{fig:scan_direct_theta}, where a majority of points are accessible by the Darwin experiment even for the smallest values of $\theta$, by opposition to the right panel where the contribution from the exotic quarks is still negligible for $\theta \sim 10^{-3}$.}

For smaller DM masses,  $m_{\tilde \xi}\lesssim 3$ TeV, DM annihilation to $\bar{\mathbb{K}} \mathbb{K}$ becomes kinematically unfavourable or impossible, and the coupling $k$ is typically small $k \ll 1$ so annihilations have to occur dominantly via $s-$channel $\tilde \nu_R$ resonance (see the left panel of Fig.~\ref{fig:scan_parameter_space_MSNUR_VS_MDM}) but also in a smaller proportion via $Z^\prime$ resonance for $m_{\tilde \xi}\sim 1-3$ TeV (as shown in  the left panel of Fig.~\ref{fig:scan_parameter_space_MZP_VS_MDM}). As most of the non SM-like fields are typically heavier than the DM mass, the quasi on-shell $\tilde \nu_R, Z^\prime$ essentially subsequently produce a pair of SM quarks or neutralinos $\tilde \chi^0_i \tilde \chi^0_i$ and charginos $\tilde \chi^+ \tilde \chi^-$ when kinematically possible. This regime offers direct detection prospects similar to the case $v_R>10$ TeV for small DM masses with a large part of the parameter space in the reach of Darwin but a non-negligible part beyond, and a subdominant below the neutrino floor. However, given the fact that the cross section for $\tilde \xi \tilde \xi \rightarrow \bar q q$ mediated by $s-$channel $\tilde \nu_R$ diagrams is velocity suppressed, this part of the parameter space offers less optimistic indirect detection prospects and might not be accessible by CTA in the near future. One can also observe a small cluster of points in the left panel of Fig.~\ref{fig:scan_direct_detection} for $m_{\tilde \xi} \simeq m_h/2 \simeq 62$ GeV, where the SM-like Higgs resonance significantly increases the annihilation cross section mediated by Higgs-diagrams and induced by scalar mixing. Interestingly, this very specific case is in the reach of the Darwin experiment.
{Another interesting feature 
is that $\tilde \nu_R$ masses are expected to be at most at around a few TeV for $m_{\tilde \xi} \lesssim 1-2$ TeV, as can be seen in the left panel of Fig.~\ref{fig:scan_parameter_space_MSNUR_VS_MDM}.}

{Let us finally point out that in the case {discussed in Eq.~(\ref{rhneutrino}),} where extra singlets of the type $\hat N, \hat S$ contribute to generate RH neutrino masses, as the cross section grows with the mass squared of the outgoing fermions (see Eq. (\ref{eq:annihilation_neutralinos})), the annihilation to RH neutrinos could become more important for larger masses, and therefore more ease to achieve the correct relic density.}

\section{Conclusions}
\label{conclusions}

We considered in this work a specific WIMP DM realization in the framework of the U$\mu\nu$SSM, which is a $U(1)'$ extension of the $\mu\nu$SSM. In order to ensure an anomaly free theory, states charged under the new gauge symmetry are introduced: exotic quarks and additional singlets under the SM gauge group. Masses for these new states are generated dynamically once the right sneutrino acquires a VEV, simultaneously generating the $\mu$-term and masses for RH neutrinos. The requirement of gauge symmetry and SUSY ensures the lightest of these new SM singlet states to be stable, and to behave as a good candidate for WIMP DM without introducing $R$-parity. This kind of DM interacts with the SM particle content via exchange of a new massive gauge boson $Z^\prime$, right sneutrinos, SM-like Higgs via scalar mixing, as well as DM exchange (see Fig.~\ref{fig:diagrams}). 

In this setup, SI (SD) DM-nucleon scatterings are mediated by Higgs via scalar 
mixing ($Z^\prime$), by interactions with light quarks within nucleons (see Fig.~\ref{fig:diagram_DD}). 
{Therefore, DM direct detection experiments can probe regions of our parameter space.}
We also pointed out that the exotic quarks offer an additional channel for SI scatterings by interacting directly with the gluons present in the nucleons and with DM by right sneutrino mediation. 
As the presence of these exotics is required by the anomaly cancellation conditions, their contribution is a rather general prediction of the U$\mu \nu$SSM. {Although it turns out to be significant only in specific corners of the parameter space of our scan range, it offers nevertheless the possibility of testing a part of the parameters in the future in the case of low values of the scalar mixing.}

Additional constraints on this scenario are imposed by $Z^\prime$ LHC searches which can exclude masses $m_{Z^\prime} \simeq 1-5$ TeV depending on the value of the $U(1)'$ gauge 
coupling (see Fig.~\ref{fig:lim1}), as well as R-hadron searches which provide a lower bound on the masses of exotic quarks of the order of the TeV scale. Concerning LHC signals of the DM particle itself, the direct production is quite suppressed because it is a SM singlet. However, in regions of the parameter space where the singlets $\tilde\xi$ are lighter than $m_{Z'}/2$, they could be produced in $Z'$ decays. The decay $Z' \to \tilde \xi_2 \tilde \xi_2$ with subsequent decay $\tilde \xi_2 \to \tilde \xi_1 \ell^+ \ell^-$ produces two pairs of collimated leptons, which can give striking signatures~\cite{Aguilar-Saavedra:2019iil}. Other decay modes such as $\tilde \xi_2 \to \tilde \xi_1 q \bar q$ are likely unobservable, as are the decays $Z' \to \tilde \xi_1 \tilde \xi_1$.

We analyzed the possibility of reproducing the observed DM relic abundance via the freeze-out mechanism in this setup, performing a numerical analysis of the viable parameter space respecting all constraints. Results from Xenon1T experiment already exclude a subdominant portion of the allowed parameter space. We identified two main regions allowed by Xenon1T, at large DM masses $m_{\tilde \xi} \gtrsim 2-3$ TeV and at smaller DM masses $ 200 ~\text{GeV}\lesssim m_{\tilde \xi} \lesssim 2-3~\text{TeV}$ (see Fig.~\ref{fig:scan_direct_detection}). 

For the case of large masses, the dynamical generation of the DM mass implies relatively large couplings with the right sneutrino. For this regime, new bosonic (right sneutrinos $\tilde \nu_R$, heavy gauge boson $Z^\prime$) and fermionic (exotic quarks $\mathbb{K}$, neutralinos $\tilde \chi_i$ and charginos $\tilde \chi^\pm$) states are the most frequent particles present in the final states of DM annihilation, and therefore essential to reproduce the correct relic abundance. For
the highest masses, $m_{\tilde \xi} \gtrsim 10^5$ GeV, the viable part of the parameter space requires couplings that are typically on the edge of perturbative unitarity. This part of the parameter space offers optimistic detection prospects as the Darwin experiment should probe the majority of the viable parameters in the following year and the remaining part should be accessible with an increased exposure. 

For the region with lower masses, achieving the correct relic abundance is less frequent as most of the annihilation channels mentioned previously are kinematicaly forbidden after imposing constraints on the new states. This regime typically relies on $s-$channel 
$\tilde\nu_R$ or $Z^\prime$ resonances with SM particles in the final states such as quarks. Relatively low couplings are typically required for such masses and therefore the direct detection prospects are less optimistic. Nevertheless, a substantial part of the parameter space will be accessible by the Darwin experiment. 

Interestingly, as many annihilation channels are usually required to achieve the correct relic abundance, non-velocity suppressed DM annihilation within large astrophysical structures could offer complementary detection prospects by indirect gamma-ray searches with the upcoming CTA.

If in the future a DM direct detection signal is reported, it is true that in principle other models could predict similar DM-nucleon scattering cross sections as the
U$\mu \nu$SSM. Nevertheless, one of the immediate predictions of our model is the simultaneous presence of exotic quarks and heavy mediators ($Z^\prime$ and $\tilde \nu_R$). Therefore, the complementary (non) observation of such states at colliders, as well as the potential indirect $\gamma$-ray signal, will be useful to 
(discard) validate the U$\mu \nu$SSM as one of the possible interpretations for the signal.


\begin{acknowledgments}

The authors would like to thank Geneviève Bélanger for useful discussions and help with the code \texttt{micrOMEGAs}.
The research of JAAS was supported by the Spanish Agencia Estatal de Investigaci\'on (AEI) through project PID2019-110058GB-C21 and by FCT project CERN/FIS-PAR/0004/2019.
The work 
of DL was supported by the Argentinian CONICET, and also acknowledges the support through PIP 11220170100154CO.
The research of CM and MP was supported by the Spanish AEI 
through the grants 
PGC2018-095161-B-I00 (EU FEDER) and IFT Centro de Excelencia Severo Ochoa SEV-2016-0597.  
MP acknowledges support by the Deutsche Forschungsgemeinschaft (DFG, German Research
Foundation) under Germany‘s Excellence Strategy – EXC 2121 “Quantum Universe” – 390833306. This work was made possible by with the support of the Institut Pascal at Université Paris-Saclay during the
Paris-Saclay Astroparticle Symposium 2021, with the support of the P2IO Laboratory of Excellence (program “Investissements d’avenir” ANR-11-IDEX-0003-01 Paris-Saclay and ANR-10-LABX-0038),
the P2I axis of the Graduate School Physics of Université Paris-Saclay,
as well as IJCLab, CEA, IPhT, APPEC, the IN2P3 master projet UCMN and EuCAPT ANR-11-IDEX-0003-01 Paris-Saclay and ANR-10-LABX-0038).
\end{acknowledgments}

\bibliographystyle{utphys}
\bibliography{munussmbib-completo_v6}

\begin{thebibliography}{100}
\bibitem{CMS:2020qrs}
{\bfseries CMS} Collaboration, ``{Search for a spin-1 heavy resonance that
  decays to a Z boson and Higgs boson in the semileptonic final states with
  Run-2 data},''.
\end{thebibliography}
\end{document}